\documentclass[aip, jcp, 11pt,reprint]{revtex4-1}
\usepackage{amsmath,amssymb}
\usepackage{graphicx}
\usepackage{multirow}
\usepackage{dcolumn}

\newcommand{\Tr}{\mathrm{Tr}}

\usepackage{braket}

\newcommand{\pder}[3][]{\frac{\partial^{#1}{#2}}{\partial{#3}^{#1}}}
\newcommand{\pders}[3]{\frac{\partial^2{#1}}{\partial{#2}\partial{#3}}}
\newcommand{\der}[2]{\frac{d{#1}}{d{#2}}}
\newcommand{\half}{{\ensuremath{\frac{1}{2}}}}
\newcommand{\thalf}{{\ensuremath{\tfrac{1}{2}}}} 
\DeclareMathOperator{\Imag}{Im}
\renewcommand{\Im}{\Imag}
\usepackage{soul} 
\usepackage{comment} 
\usepackage{dcolumn}
\newcolumntype{d}[1]{D{.}{.}{#1}}

\begin{document}
\title{Semiclassical analysis of the quantum instanton approximation}
\author{Christophe L. Vaillant}
\email{christophe.vaillant@epfl.ch}
\affiliation{Laboratory of Theoretical Physical Chemistry, Institut des Sciences et Ing\'enierie Chimiques, \'Ecole Polytechnique F\'ed\'erale de Lausanne (EPFL), CH-1015, Lausanne, Switzerland}
\author{Manish J. Thapa}
\affiliation{Laboratory of Physical Chemistry, ETH Z\"urich, 8093 Z\"urich, Switzerland}
\author{Ji\v{r}\'{\i} Van\'{\i}\v{c}ek}
\email{jiri.vanicek@epfl.ch}
\affiliation{Laboratory of Theoretical Physical Chemistry, Institut des Sciences et Ing\'enierie Chimiques, \'Ecole Polytechnique F\'ed\'erale de Lausanne (EPFL), CH-1015, Lausanne, Switzerland}
\author{Jeremy O. Richardson}
\email{jeremy.richardson@phys.chem.ethz.ch}
\affiliation{Laboratory of Physical Chemistry, ETH Z\"urich, 8093 Z\"urich, Switzerland}
\date{\today}
\begin{abstract}
We explore the relation between the quantum and semiclassical instanton approximations for the reaction rate constant.
From the quantum instanton expression, we analyze the contributions to the rate constant in terms of minimum-action paths and find that two such paths dominate the expression.
For symmetric barriers, these two paths join together to describe the semiclassical instanton periodic orbit.
However, for asymmetric barriers, one of the two paths takes an unphysically low energy and
dominates the expression, leading to order-of-magnitude errors in the rate predictions.
Nevertheless, semiclassical instanton theory remains accurate.
We conclude that semiclassical instanton theory can only be obtained directly
from the semiclassical limit of the quantum instanton for symmetric systems.
We suggest a modification of the quantum instanton approach which avoids sampling the spurious path
and thus has a stronger connection to semiclassical instanton theory, giving numerically accurate predictions even for very asymmetric systems in the low temperature limit.
\end{abstract}
\maketitle

\section{Introduction}
The inclusion of nuclear quantum effects in molecular dynamics
calculations is a challenging task. One of the main difficulties in
calculating exact quantum dynamics is the need to solve the time-dependent
Schr\"odinger equation with many degrees of freedom. For atomistic descriptions of
most molecular systems the dynamics are (currently)
computationally impossible to solve exactly, due to the
exponentially-increasing size of the basis sets required for these
calculations.
For this reason, approximate quantum dynamics methods based on path integrals are increasingly favoured.

A rigorous formulation of reaction rate theory is provided by
the flux-side or flux-flux time correlation function formalism.\cite{Miller1983}
These correlation functions can be evaluated with approximate quantum-dynamics methods,
including linearized
semiclassical initial-value representation (LSC-IVR),\cite{Miller2001}
centroid molecular dynamics,\cite{Cao1994,Jang1999} ring-polymer
molecular dynamics (RPMD),\cite{Craig2004,Craig2005a,Craig2005b} and
the recently-developed Matsubara dynamics\cite{Hele2015a,Hele2015b}
and its approximations.\cite{Willatt2018,Trenins2018}
The main approximation made by these methods is to neglect quantum coherences,
which is not expected to be a cause of significant error
for complex systems in thermodynamic equilibrium.\cite{Miller2001SCIVR} Thus, solving the system exactly to obtain the long-time
dynamics is in many cases unnecessary as the relevant processes can be approximated
to a good accuracy at short times. In fact, if the problem is well defined, no time propagation is necessary at all and one can turn to
so-called quantum transition state theories (QTSTs), where the expressions involving time correlation functions are approximated using only their properties at zero time.\cite{Miller1974, Pollak1998}

Semiclassical instanton (SCI) theory provides one of the simplest formulations of a QTST for multidimensional
tunneling.\cite{Miller1975,Perspective} It can be derived rigorously by taking asymptotic ($\hbar\rightarrow0$) approximations to the flux-flux correlation function\cite{Richardson2016} and is known in a number of formulations,\cite{InstReview} including one obtained from the ``ImF'' approximation,\cite{Affleck1981} which can all be shown to be equivalent.\cite{Althorpe2011}
SCI theory is defined in terms of the dominant tunneling pathway located on the full-dimensional potential-energy surface, and employs a harmonic approximation for fluctuations around this pathway. The ring-polymer instanton approach\cite{Andersson2009Hmethane,Richardson2009,Rommel2011locating} provides a computationally efficient algorithm for applying the theory in practice, and can be automated for application to complex systems.\cite{Vaillant2019}

Although SCI theory is very powerful, the harmonic expansion about a single dominant tunneling path can introduce errors, especially for low-frequency anharmonic systems. An alternative method, known as the quantum instanton (QI),\cite{Miller2003} was developed as an attempt to correct the quantitative deficiencies of the semiclassical approximation. This method goes beyond the harmonic approximation employed by SCI theory and samples paths using efficient path-integral Monte Carlo or molecular dynamics methods.\cite{Yamamoto2004} The QI method has been used to calculate reaction rates in many multidimensional systems,\cite{Zhao2004, Wang2009,Yamamoto2005} and is a particularly efficient approach for the direct calculation of kinetic isotope effects.\cite{Vanicek2005,Karandashev2017b,Karandashev2015,Karandashev2017a} However, for large asymmetries (meaning the reaction is highly exothermic or endothermic) the rate constant predicted by the QI method has been observed to give large errors, whereas SCI theory remains well-behaved.\cite{Karandashev2017c,InstReview} 

There are also a number of other path-integral-based QTSTs developed to go beyond the SCI approximation. Unlike in classical mechanics,\cite{Chandler1978TST} it is not possible to derive them directly from the flux-side correlation function, as this function tends to zero at zero time if used in its standard form. Instead, QTSTs have been proposed based on a connection to semiclassical instanton theory, including a free-energy version of instanton theory\cite{Mills1997QTST} and ring-polymer transition-state theory, which is itself related to the RPMD rate theory.\cite{Richardson2009} More recently, ring-polymer transition-state theory has been rederived from a generalized flux-side correlation function which yields a good approximation to the rate at zero time.\cite{Hele2013,Althorpe2013} This provides an extension of the centroid-based QTST of  Voth, Chandler and Miller,\cite{Voth+Chandler+Miller}
a method which does not dominantly sample the instanton and thus fails for asymmetric barriers.\cite{Makarov1995QTST}

In summary, it appears that the QTSTs which work best are those with the strongest connection to the SCI. In this article, we examine the relation between the QI and SCI theories, in order to investigate the very plausible conjecture that the SCI is a semiclassical/steepest-descent approximation to the QI; put another way, that the QI method is an improvement on the SCI with more accurate sampling of the paths. Surprisingly, we find that the two methods are not directly connected to each other except in the case of symmetric systems. We perform a semiclassical analysis of the quantities used in the QI method
and find that the dominant paths that contribute are not the same as those which define the instanton periodic orbit. This leads to the observed error of the QI method for systems with large asymmetries and low temperatures. We then suggest an approach for modifying the QI expression to enforce sampling of the instanton orbit, which is seen to greatly improve the results.

The paper is organized as follows. We first give a derivation of the QI method from first principles in Sec.~\ref{section:QI}, followed by a short investigation of the breakdown of the predicted rate for asymmetric systems. Motivated by this breakdown, we analyze the contributions from semiclassical paths in Sec.~\ref{section:SCI}, and we show that the SCI expression can only be recovered if the spurious paths are removed. We suggest adding a projection operator to solve this problem and describe an improved approach, which we call the projected quantum instanton method, in Sec.~\ref{section:projected}. We follow this with a discussion of the numerical results of the various methods in Sec.~\ref{section:discussion}, including the semiclassical approximations, which
justifies the analysis in terms of semiclassical paths, before concluding in Sec.~\ref{section:conclusions}.

\section{Quantum instanton approximation}
\label{section:QI}
In the following section we re-examine the derivation of the QI method for calculating reaction rates. The derivation is based on a saddle-point approximation to the time integral of the flux-flux correlation function, and we show that the method breaks down for asymmetric barriers at low temperatures. The goal is to have a self-contained discussion of the QI method before undertaking a more detailed semiclassical analysis. There are two versions of the QI method: the original, which we shall continue to call QI, and a variant, which we call the second-order cumulant expansion (2OCE).

Although we limit ourselves to one dimension throughout the paper for simplicity, multidimensional extensions of the QI method exist.\cite{Miller2003,Yamamoto2004,Zhao2004,Vanicek2005}
Our conclusions remain valid for these multidimensional cases.

\subsection{Derivation}
We begin by rederiving the QI approximation from the Miller--Schwartz--Tromp expression for the reaction rate constant\cite{Miller1983}
\begin{equation}
k Q_\mathrm{r} = \frac{1}{2} \int^{\infty}_{-\infty} \! \! \! dt \; C_\mathrm{ff} (t).
\label{1d:reactionrate}
\end{equation}
Here, $Q_\mathrm{r}$ is the reactant partition function and $C_\mathrm{ff}$ is the symmetrized flux-flux time correlation function,
\begin{equation}
C_\mathrm{ff} (t) = \Tr \left( e^{-\beta \hat{H}/2} \hat{F}_1 e^{-\beta \hat{H}/2} e^{i \hat{H} t/\hbar} \hat{F}_2 e^{-i \hat{H} t/\hbar} \right),
\label{1d:symtcf}
\end{equation}
generalized to two dividing surfaces. 
The Hamiltonian for a particle in a one-dimensional potential, $V(x)$, is $\hat{H} = \hat{p}^2/2m + V(\hat{x})$ and
the flux operators are given by
\begin{equation}
\hat{F}_{j}= \frac{1}{2m} \left( \hat{\delta}_{j}\, \hat{p} + \hat{p}\,\hat{\delta}_{j} \right) \qquad (j\in\{1,2\}),
  \label{1d:fluxoperator}
\end{equation}
where $m$ is the mass and $\hat{\delta}_j= \delta (\hat{x} - x_j)$ indicates a Dirac delta function centred at the $j$-th dividing surface located at $x = x_j$. Because both $\hat{H}$ and $\hat{F}_{j}$ are Hermitian as well as real operators, the correlation function [Eq.~(\ref{1d:symtcf})] is a real and even function of $t$.

The main idea used in deriving the QI expression is to approximate the time integral in Eq.~\eqref{1d:reactionrate} using a steepest-descent (saddle-point) approximation taken around $t=0$,
\begin{equation}
\int^{\infty}_{-\infty} \! \! dt \;  e^{f(t)} \approx \int^{\infty}_{-\infty} \! \! dt \;  e^{f(0)+\ddot{f}(0)t^2/2} 
= \sqrt{\frac{2\pi}{-\ddot{f}(0)}} e^{f(0)} ,
\label{1d:steepestdescent}
\end{equation}
where the dot implies differentiation with respect to time. In order to use this approximation, it is necessary that $\dot{f}(0)=0$, which is guaranteed for even functions of $t$, and also that $\ddot{f}(0)<0$. This steepest-descent approximation makes it possible to express the full reaction rate in terms of expressions defined at $t=0$, which can then be evaluated exactly with imaginary-time path-integral methods.
The original QI expression [see Eq.~(\ref{1d:quantuminstantonsp}) below] has been derived with one of two approaches: one which uses the energy-integral formulation of the reaction rate (with complicated transformations and two separate steepest-descent integrations)\cite{Miller2003} and another which involves multiplying and dividing by the delta-delta correlation function, 
\begin{equation}
C_\mathrm{dd} (t) = \Tr \left( e^{-\beta \hat{H}/2} \hat{\delta}_1 e^{-\beta \hat{H}/2} e^{i \hat{H} t/\hbar} \hat{\delta}_2 e^{-i \hat{H} t/\hbar}\right),
\label{1d:cdd}
\end{equation}
as the first step.\cite{Vanicek2005,Aieta2017} Both derivations use the same prescription for choosing the dividing surfaces $x_1$ and $x_2$.

Evaluating the trace in Eq.~\eqref{1d:symtcf} in the position representation, we find the well-known expression\cite{Miller1983,Aieta2017}
\begin{equation}
C_\mathrm{ff} (t) = \left( \frac{\hbar}{2m} \right)^2 \left( \frac{\partial^2 \rho}{\partial x_1 \partial x_2} \rho^\ast  - \frac{\partial \rho}{\partial x_1}\frac{\partial \rho^\ast}{\partial x_2} \right) + \mathrm{c.c.},
\label{1d:cff}
\end{equation}
where ``c.c.'' denotes the complex conjugate, and
\begin{equation}
\rho \equiv \rho(x_1, x_2, t) = \langle x_2 | e^{-\beta \hat{H}/2 - i \hat{H} t/\hbar} | x_1 \rangle.
\label{1d:rho12}
\end{equation}
The delta-delta correlation function [Eq.~(\ref{1d:cff})] is then simply
\begin{equation}
C_\mathrm{dd}(t) = |\rho(x_1,x_2,t)|^2.
\label{1d:deltadelta}
\end{equation}
For brevity, we will drop the explicit dependence of the time correlation functions on the dividing surfaces in our notation. To apply the steepest-descent approximation [Eq.~\eqref{1d:steepestdescent}] to Eq.~\eqref{1d:reactionrate}, we set
\begin{equation}
f_\mathrm{ff}(t) = \ln  C_\mathrm{ff}(t) ,
\label{1d:cffsteepestfunction}
\end{equation}
which we note obeys the stationarity condition at time $t=0$ because $\dot{C}_\mathrm{ff}(0)=0$. We are still free to choose the locations of the dividing surfaces, $(x_1,x_2)$, and we should do this in such a way to ensure that $C_\mathrm{ff}(t)$ has a maximum at $t=0$. Many choices for the dividing surfaces have been suggested, but some choices lead to a local minimum at zero time for $C_\mathrm{ff}(t)$, rather than a maximum.\cite{Aieta2017} The original and most common choice is that the dividing surfaces obey
\begin{equation}
\pder{\rho(x_1,x_2,0)}{x_1} = \pder{\rho(x_1,x_2,0)}{x_2}=0,
\label{1d:divsurfaces}
\end{equation}
which typically leads to $C_\mathrm{ff}(t)$ being a maximum at $t=0$. In our experience, there is always a solution to these equations with $x_1=x_2$, known as merged dividing surfaces, which describe either a minimum or first-order saddle point of $\rho(x_1,x_2,0)$.
In the former case, another set of solutions exists as saddle points with $x_1\ne x_2$, leading to what is known as split dividing surfaces. Adopting one of these choices of dividing surfaces such that $x_1$ and $x_2$ satisfy Eq.~\eqref{1d:divsurfaces}, we note that $C_\mathrm{ff}(0)=C^\mathrm{sp}_\mathrm{ff}(0)$, where
\begin{equation}
C^\mathrm{sp}_\mathrm{ff}(t) = \left( \frac{\hbar}{2m} \right)^2 \frac{\partial^2}{\partial x_1 \partial x_2} C_\mathrm{dd}(t) ,
\label{1d:stationarycff}
\end{equation}
and thus suggest the approximation $C_\mathrm{ff}(t) \approx C^\mathrm{sp}_\mathrm{ff}(t)$ for $t\ne0$.

We are now faced with a choice for performing the steepest-descent integration in time: we could either perform this approximate time integral on $C_\mathrm{dd}(t)$ and then take the spatial derivatives, or we could perform the integral directly on $C_\mathrm{ff}(t)$.
Choosing first to perform the integral over time before evaluating the derivatives with respect to the positions of surfaces, the new function for the steepest-descent approximation and its second time derivative (in terms of the energy variance $\Delta H_\mathrm{dd}^2$) are
\begin{subequations}
\begin{align}
f_\mathrm{dd}(t) &= \ln  C_{\mathrm{dd}}(t) \\
\left.\frac{d^2 f_\mathrm{dd}}{d t^2}\right\rvert_{t=0} &= \frac{\ddot{C}_{\mathrm{dd}}(0)}{C_{\mathrm{dd}}(0)} = -\frac{2}{\hbar^2} \Delta H_\mathrm{dd}^2,
\end{align}
\label{1d:stationaryfunctiondd}%
\end{subequations}
where $\dot{C}_\mathrm{dd}(0)=0$ and
\begin{equation}
\ddot{C}_\mathrm{dd}(t) = 2|\dot{\rho}|^2 + \ddot{\rho}\rho^\ast +  \rho \ddot{\rho}^\ast.
\label{1d:cddddot}
\end{equation}
The QI rate is then defined as
\begin{equation}
\begin{split}
k_\mathrm{QI} Q_\mathrm{r} &= \half \left( \frac{\hbar}{2m} \right)^2 \frac{\partial^2 C_\mathrm{dd} (0) }{\partial x_1 \partial x_2}  \int_{-\infty}^\infty \! dt \, e^{-\Delta H_\mathrm{dd}^2t^2/\hbar^2}\\
&=\frac{\hbar \sqrt{\pi}}{2 \Delta H_\mathrm{dd}} C^\mathrm{sp}_\mathrm{ff}(0),
\end{split}
\label{1d:quantuminstantonsp}%
\end{equation}
where we have assumed that the spatial derivatives of $\Delta H_\textrm{dd}$ can be neglected as they are much smaller than the derivatives of $C_\mathrm{dd}(0)$. Provided that the ratio $C_\mathrm{ff}(t)/C_\mathrm{dd}(t)$ is slowly-varying in time, Eq.~\eqref{1d:quantuminstantonsp} can be generalized for dividing surfaces that do not obey the condition in Eq.~\eqref{1d:divsurfaces}, such that\cite{Aieta2017}
\begin{equation}
k_\mathrm{QI} Q_\mathrm{r} = \frac{\hbar \sqrt{\pi}}{2 \Delta H_\mathrm{dd}} C_\mathrm{ff}(0),
\label{1d:quantuminstanton}
\end{equation}
where we have slightly abused the notation and kept the QI subscript in this final expression.

An alternative approach is derived by employing a saddle-point approximation directly to the time integral of $C_\mathrm{ff}(t)$. In order to use this approximation, there is no particular requirement on the choice of dividing surfaces,
provided $C_\mathrm{ff}(t)$ is a maximum at $t=0$ for that choice.
The idea of this direct integration has previously been introduced in the context of creating a ``cumulant expansion'', where the expansion is truncated at the second order term \cite{Ceotto2005} (although it could be extended to higher order terms
and, in addition, the exact high-temperature behavior could be incorporated analytically).\cite{Ceotto2005, Yang2006, Predescu2004}
The steepest-descent function is then
\begin{subequations}
\begin{align}
f_\mathrm{ff}(t) &= \ln  C_{\mathrm{ff}}(t) \\
\left.\frac{d^2 f_\mathrm{ff}}{d t^2}\right|_{t=0}&= \frac{\ddot{C}_{\mathrm{ff}}(0)}{C_{\mathrm{ff}}(0)} = -\frac{2}{\hbar^2} \Delta H_\mathrm{ff}^2,
\end{align}
\label{1d:stationaryfunctionff}%
\end{subequations}
where
\begin{equation}
\begin{split}
\ddot{C}_\mathrm{ff}(t) = & \left( \frac{\hbar}{2m} \right)^2 \left( \frac{\partial^2 \rho}{\partial x_1 \partial x_2} \ddot{\rho}^\ast+ 2  \frac{\partial^2 \dot{\rho}}{\partial x_1 \partial x_2}\dot{\rho}^\ast +  \frac{\partial^2 \ddot{\rho}}{\partial x_1 \partial x_2}\rho^\ast \right.\\
 &\left. - \frac{\partial \ddot{\rho}}{\partial x_1} \frac{\partial \rho^\ast}{\partial x_2} -2 \frac{\partial \dot{\rho}}{\partial x_1} \frac{\partial \dot{\rho}^\ast}{\partial x_2}-\frac{\partial \rho}{\partial x_1} \frac{\partial \ddot{\rho}^\ast}{\partial x_2} \right)+  \mathrm{c.c.},
\end{split}
\label{1d:cffddot}
\end{equation}
which defines the energy variance $\Delta H_\mathrm{ff}^2$. The second-order cumulant expansion (2OCE) expression that results from this procedure is very similar to Eq.~\eqref{1d:quantuminstanton}, namely
\begin{equation}
\begin{split}
k_\mathrm{2OCE} Q_\mathrm{r} &= \half C_\mathrm{ff} (0) \int_{-\infty}^\infty \! dt \, e^{-\Delta H_\mathrm{ff}^2t^2/\hbar^2}\\
&= \frac{\hbar \sqrt{\pi}}{2 \Delta H_\mathrm{ff}} C_\mathrm{ff}(0).
\end{split}
\label{1d:2oce}
\end{equation}
We will show that both of these quantum instanton approximations have similar properties and break down for asymmetric systems for similar reasons.

\subsection{Behaviour of the quantum instanton method for asymmetric systems}
To test the approximations, it is useful to apply the QI and 2OCE methods to a simple 1D Eckart barrier,
\begin{equation}
V(x) = \frac{(1 -\alpha) V_0}{1 + e^{-2ax}} + \frac{ (1 + \sqrt{\alpha})^2 V_0}{4 \cosh^2(ax)},
\label{asymmetriceckart}
\end{equation}
where, choosing the same parameters as in Ref.~\onlinecite{Miller2003}, $V_0= 0.425$~eV, $a= 1.36 \; a_0^{-1}$, the mass is 1060 $m_\mathrm{e}$
($a_0$ is the Bohr radius and $m_\mathrm{e}$ denotes the electron mass). $\alpha$ is the dimensionless asymmetry parameter controlling the degree of exothermicity between the reactants and products such that the barrier is symmetric for $\alpha=1$. We investigated a variety of different temperatures and asymmetries, and found that the QI method shows pathological behaviour for low temperatures and large asymmetries. We will demonstrate this behaviour throughout the paper, unless otherwise stated, with a particular choice of $T=100$~K and $\alpha=1.425$, for which the QI prediction is an order of magnitude too large compared to the exact rate.

\begin{figure}[t]
\centering
\includegraphics{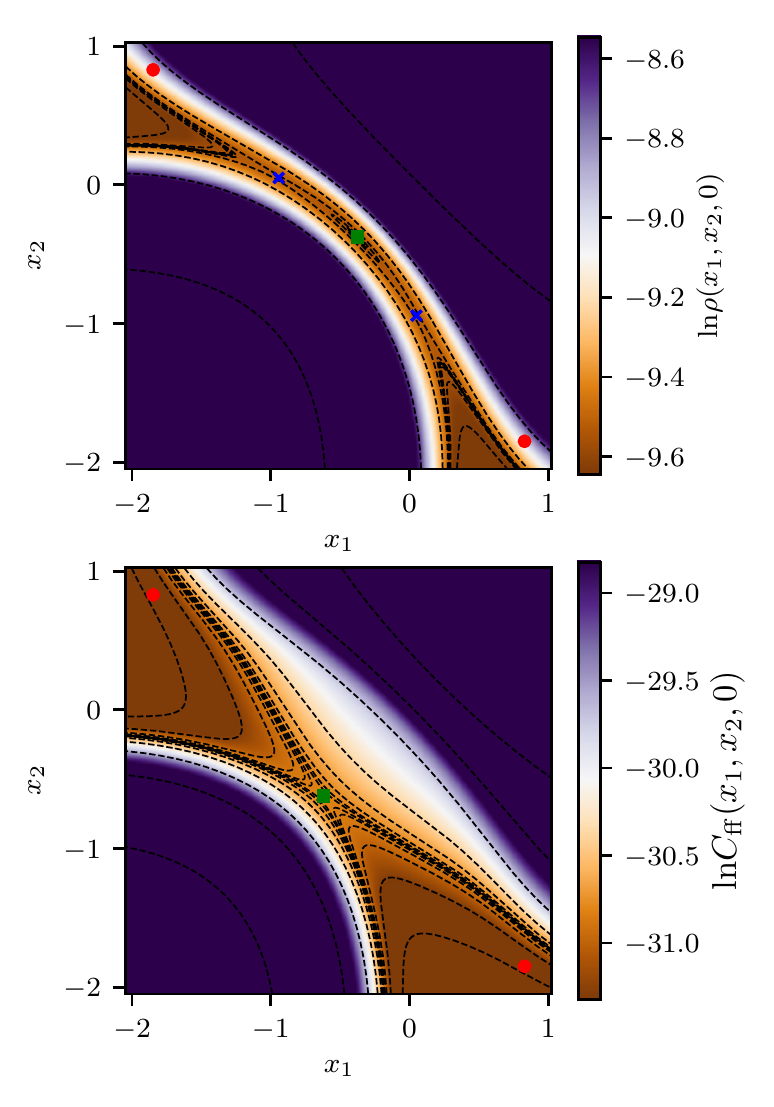}
\caption{\label{asymmetricrho12} Contour plots of $\rho(x_1,x_2,0)$ and $C_\mathrm{ff}(0)$ for the asymmetric barrier at 100 K and $\alpha= 1.425$, plotted as functions of the dividing surface locations. The saddle points of $\rho$ (and therefore $C_\mathrm{dd}(0)$, which is simply the square of $\rho$) which define the split surfaces are shown as blue crosses, the semiclassical instanton turning points are shown as red circles, and the merged surfaces locations of either $C_\text{dd}(0)$ or $C_\text{ff}(0)$ are shown by the green squares. These coordinates are given in Table~\ref{tab:surfaces}. All quantities are in atomic units.}
\end{figure}

It would be reasonable to think that the QI method could be improved if a more suitable choice of dividing surfaces were found. We will therefore consider a range of choices for the dividing surfaces, listed in Table~\ref{tab:surfaces}, with the above choice of parameters, showing examples of both split saddle points and a merged minimum of $C_\mathrm{dd}(0)$ and a merged saddle point of $C_\mathrm{ff}(0)$. \footnote{We were unable to locate an asymmetric system with split saddle points of $C_\mathrm{ff}(0)$.}  The positions of the dividing surfaces are also indicated in Fig.~\ref{asymmetricrho12}, showing the behavior of the propagator and the flux-flux correlation function as functions of the dividing surface locations. It is possible to choose dividing surfaces that are not at special points,\cite{Aieta2017} but we will show that it is not generally possible to find dividing surfaces that fix all the problems of the QI and 2OCE methods without needing to evaluate quantities at $t\ne0$.

\begin{table}[t]
  \caption{\label{tab:surfaces}Definition and positions of dividing surfaces for the asymmetric Eckart barrier with $\alpha=1.425$ and $T=100$ K. 
  The remaining parameters are defined in the main text.}
  \begin{ruledtabular}
    \begin{tabular}{lcd{2.3}d{2.3}}
      Choice of surfaces & Shorthand & \multicolumn{1}{c}{$x_1$} & \multicolumn{1}{c}{$x_2$}\\
      \hline
      Split saddle points of $C_{\mathrm{dd}}$ & A &-0.968&0.063\\
      Merged minimum of $C_{\mathrm{dd}}$& B & -0.379 & -0.379\\
      Merged saddle point of $C_{\mathrm{ff}}$ & C & -0.623 & -0.623\\
      Semiclassical turning points & D & -1.847 & 0.828\\
    \end{tabular}
  \end{ruledtabular}
\end{table}

\begin{figure}[t]
\centering
\includegraphics{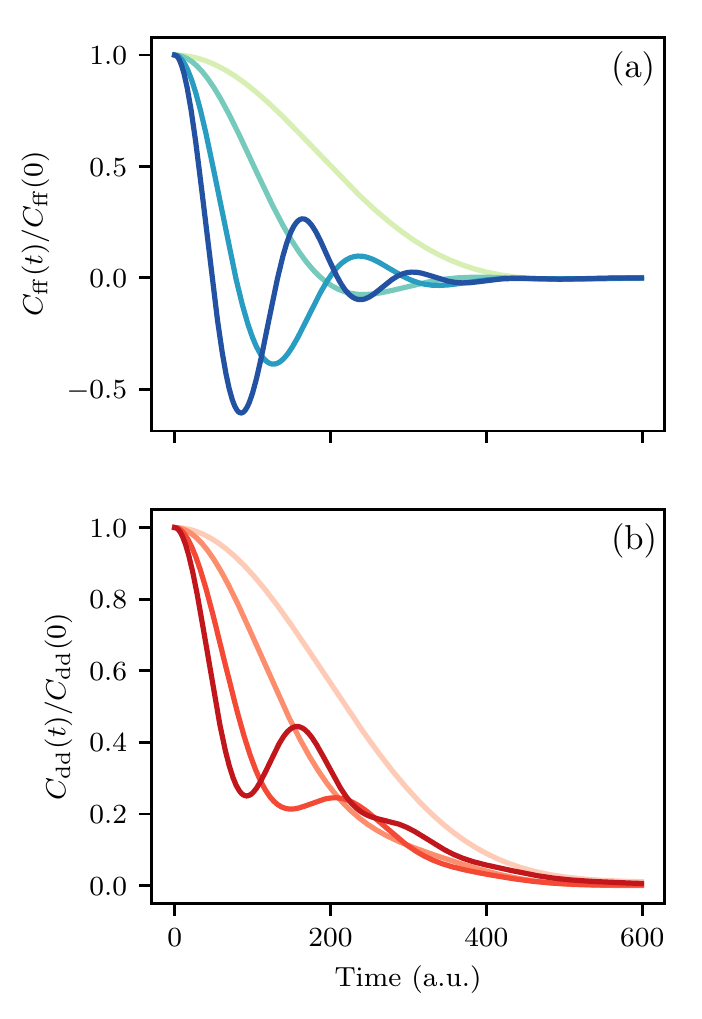}
\caption{\label{asymmetrictcfs} (a) Flux-flux and (b) delta-delta time correlation functions as functions of time for several values of the asymmetry parameter $\alpha$. Darker lines indicate larger values of $\alpha$, with $\alpha=$ 1, 2, 3, and 4.
For each value of $\alpha$, the dividing surfaces are reoptimized  with the split saddle points on $C_\mathrm{dd}$ chosen if they exist, corresponding to A-type surfaces in Table~\ref{tab:surfaces}. In this way, the same surfaces are used for both $C_\mathrm{ff}$ and $C_\mathrm{dd}$, although similar behaviour is also seen for other choices of dividing surfaces.
}
\end{figure}

To summarize the previous section, the QI and 2OCE methods essentially approximate the time dependence of
the relevant time correlation functions as Gaussians. The functions $C_\mathrm{dd}(t)$ and $C_\mathrm{ff}(t)$ are shown in Fig.~\ref{asymmetrictcfs}, where the surfaces are chosen to be the split surfaces  of $\rho$ if they exist, or the merged surfaces of $\rho$ otherwise. For increasing asymmetry $\alpha$ it is clear that the functions become less and less Gaussian (a similar situation to that discussed in Refs.~\onlinecite{Huo2013PLDM,nonoscillatory} for electron-transfer rates). This will lead to a significant error in the prediction of the rate constant. In fact, we find that none of the choices considered for the dividing surfaces make much difference in the QI prediction for the rate, as the QI prediction still deviates significantly from the exact value. In order to gain insight into this behaviour, in the next section we analyze the correlation functions in terms of semiclassical paths and discover the cause of the breakdown of the approximation.

\section{Semiclassical analysis}
\label{section:SCI}
In this section, we will examine the various terms in the QI and 2OCE approximations using a semiclassical analysis,
which gives the asymptotic behaviour in the $\hbar\rightarrow0$ limit of a quantum-mechanical expression in terms of minimum-action paths. We will identify the dominant contributions that cause the observed deviation from the semiclassical instanton expression and determine what improvements would be necessary to fix these problems.

\subsection{Semiclassical contributions to quantum instanton quantities}
All terms in the QI and 2OCE rate expressions are defined at $t=0$ and thus all of our analysis can be performed
using the imaginary-time quantum propagator, $\rho(x_1,x_2,0)$, and its derivatives. Its semiclassical limit is the imaginary-time van Vleck propagator, which involves a sum over all minimum-action paths which travel from $x_1$ to $x_2$ in imaginary time $\beta\hbar/2$. These are clearly the paths which dominate in a path-integral Monte Carlo evaluation of the quantities, in which paths are weighted by the exponential of minus the action. The minimum-action paths follow imaginary-time classical trajectories, which are equivalent to Newtonian trajectories on the upside-down potential-energy surface.\cite{Miller1971density}
In fact, there are two such minimum-action paths, one which bounces to the left, and one to the right. Note that the imaginary-time classical trajectories which travel directly from one dividing surface to the other or which bounce more than once are saddle points of the action \cite{Gutzwiller1990} and do not contribute to the semiclassical limit.
The semiclassical limit of the quantum propagator is thus given by
\begin{equation}
	\rho = \braket{x_2 | e^{-\beta\hat{H}/2} | x_1}\underset{\hbar\rightarrow 0}{\thicksim} K_\ell + K_r,
\label{sc:rho}
\end{equation}
where  $\underset{\hbar\rightarrow 0}{\thicksim}$ denotes the asymptotic behaviour in the limit $\hbar \rightarrow 0$,\cite{BenderBook} while $K_\ell$ and $K_r$ are the semiclassical contributions from the left and right paths, given by
  \begin{equation}
     K_\gamma = (2\pi\hbar)^{-\half} \, \left(-\pders{S_\gamma}{x_1}{x_2}\right)^\half e^{-S_\gamma/\hbar},
       \label{sc:propagators}
  \end{equation}
where $\gamma\in\{\ell,r\}$.
These propagators are functions of $x_1,x_2$ and $\tau_\gamma$, and in this case both have the same imaginary-time length $\tau_\gamma=\beta\hbar/2$.
The paths are written in terms of the classical Euclidean action (the action in imaginary time)
\begin{equation}
  S_\gamma = \int_0^{\tau_\gamma} \! d\tau' \left\lbrace \frac{1}{2} m \left\lvert \frac{d X_\gamma(\tau')}{d \tau'} \right\rvert^2 + V[X_\gamma(\tau')] \right\rbrace,
\label{sc:actions}
\end{equation}
where $X_\gamma(\tau')$ are the imaginary-time-dependent left and right paths from $x_1$ to $x_2$ via a ``bounce''.\cite{Richardson2015} Each action is related to an eikonal $W_\gamma$ (also known as the reduced action) by the Legendre transform \cite{Gutzwiller1990,InstReview}
\begin{equation}
  S_\gamma \equiv S_\gamma(x_1,x_2,\tau_\gamma) = W_\gamma(x_1,x_2,E_\gamma) + E_\gamma\tau_\gamma,
  \label{sc:legendretransform}
\end{equation}
and
\begin{equation}
  W_\gamma \equiv W_\gamma(x_1,x_2,E_\gamma) = \int_{X_\gamma}\! dx \, \sqrt{2m[V(x)-E_\gamma]} ,
  \label{sc:eikonal}
  \end{equation}
where $x$ is integrated along the relevant path $X_\gamma$. The energy, $E_\gamma$, is chosen to solve $\pder{W_\gamma}{E_\gamma}=-\tau_\gamma$ and the path bounces at the point where $E_\gamma=V(x)$.

\begin{figure}[!t]
	\includegraphics{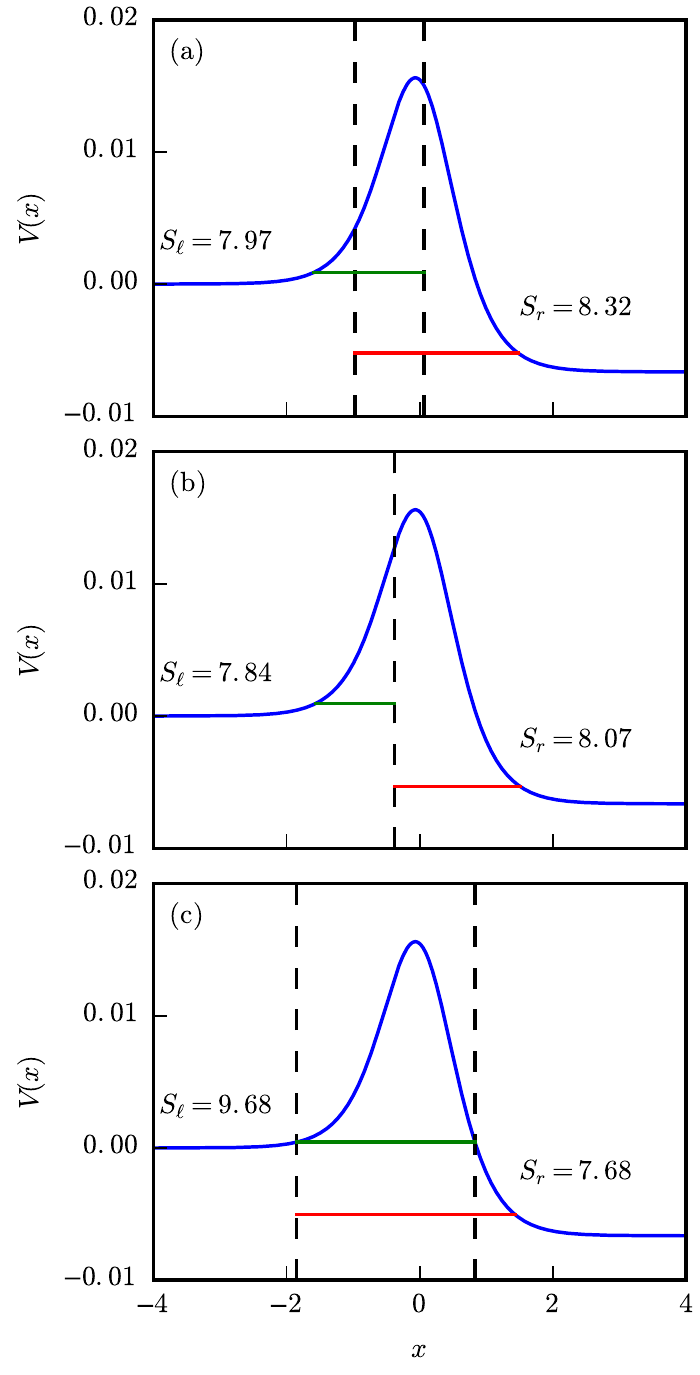}
	\caption{\label{fig:sc_paths} The two semiclassical minimum-action trajectories are shown plotted at their values of the energy for three different choices of dividing surfaces (indicated by dashed lines). The green trajectory starts at one of the dividing surfaces with initial momentum in the negative direction, bounces against the barrier on the left and returns to the other dividing surface with positive momentum. The red trajectory travels in the opposite direction and bounces against the right-hand side of the barrier. The dividing surfaces are chosen at (a) the saddle points of the quantum-mechanical $\rho(x_1,x_2,0)$, (b) the merged-surface minimum of $\rho(x_1,x_2,0)$, and (c) the semiclassical turning points. These choices correspond to A, B, and D in Table \ref{tab:surfaces}, respectively.  The choice C gives qualitatively the same picture as B. The actions of each semiclassical path are also indicated. In (c), the green line follows one half of the instanton periodic orbit. All quantities are given in atomic units.}
\end{figure}

The two minimum-action paths are shown for the A, B, and D dividing surfaces in Fig.~\ref{fig:sc_paths}. It is important to notice that the two paths do not join together to form the semiclassical instanton solution, which is an imaginary-time periodic orbit with constant energy. Because the reaction is exothermic, for all choices of dividing surfaces the left-bouncing path (green) follows half of the instanton periodic orbit but the right-bouncing path (red) is spurious and has a much lower energy, $E_r$. This remains true even when using dividing surfaces placed at the turning points of the instanton trajectory (D), as advocated in the appendix of Ref.~\onlinecite{Miller2003}. In all these cases, the second half of the instanton periodic orbit cut at the location of the dividing surfaces is a first-order saddle point of the action, as it passes through a conjugate point,\cite{Gutzwiller1990} which is why the minimum-action path must follow a different classical trajectory. In fact, we found that for this system below about 142 K there is no way to split the instanton into two trajectories of equal imaginary time without encountering a conjugate point. The picture which we show is inconsistent with the qualitative picture suggested
in the appendix of the original QI paper\cite{Miller2003} and, as we will show, the spurious path is the crux of the observed error of the QI method in asymmetric systems.

We use this semiclassical analysis in terms of minimum-action paths to find the dominant contributions to the various quantities used in the QI and 2OCE approximations.
For instance, the semiclassical limit of $C_\mathrm{dd}(0)$ is simply
\begin{equation}
  C_\mathrm{dd}(0) \underset{\hbar\rightarrow 0}{\thicksim} (K_\ell + K_r)^2 = K_\ell^2 + 2K_\ell K_r + K_r^2.
  \label{sc:cdd}
\end{equation}

The semiclassical limits of the spatial first derivatives and mixed second derivative of the imaginary-time propagator are
\begin{subequations}
\begin{align}
	\pder{\rho}{x_1} & \underset{\hbar\rightarrow 0}{\thicksim} \frac{1}{\hbar} \left(-p_1^\ell K_\ell + p_1^r  K_r \right),
	\\
	\pder{\rho}{x_2} & \underset{\hbar\rightarrow 0}{\thicksim} \frac{1}{\hbar} \left(-p_2^\ell K_\ell + p_2^r  K_r \right),
	\\
	\frac{\partial^2 \rho}{\partial x_1 \partial x_2} & \underset{\hbar\rightarrow 0}{\thicksim} \frac{1}{\hbar^2} (p_1^\ell p_2^\ell K_\ell + p_1^r p_2^r K_r),
\end{align}
 \label{SCdrhodx}%
\end{subequations}
where $p_j^\gamma = +\sqrt{2m[V(x_j)-E_\gamma]}$ for $j \in \{1,2\}$. Note that we take the positive root such that $p_j^\gamma$ is always a positive scalar and is therefore the magnitude of the momentum (without the direction). It is also clear from Fig.~\ref{fig:sc_paths} that, contrary to what was previously thought,\cite{Miller2003, Vanicek2005} the dividing surfaces according to choice A in Table~\ref{tab:surfaces}
are not close to the turning points of the semiclassical instanton,
and under a semiclassical analysis there is no relation between the two choices.

Using Eqs.~\eqref{1d:cff}, \eqref{sc:rho} and \eqref{SCdrhodx}, we find the semiclassical limit of $C_\mathrm{ff}(0)$ to be
\begin{equation}
	C_\mathrm{ff}(0) \underset{\hbar\rightarrow 0}{\thicksim} \frac{1}{2m^2} 
	(p_1^\ell + p_1^r) (p_2^\ell + p_2^r) K_\ell K_r.
    \label{sc:semiclassicalcff}
\end{equation}
It is important to note that the terms proportional to $K_\ell^2$ and $K_r^2$, which appear in $C_\mathrm{dd}(0)$, completely cancel out in the semiclassical limit of $C_\mathrm{ff}(0)$ as a consequence of the flux operators.

The real time derivatives (denoted by dots above quantities, as above) of the semiclassical propagator can be written using
the Cauchy-Riemann equations, 
$\dot{K}_{\gamma}=i \pder{K_{\gamma}}{\tau_\gamma}$ and $\dot{E}_{\gamma}=i \pder{E_{\gamma}}{\tau_\gamma}$, as
\begin{equation}
    \dot{\rho} \underset{\hbar\rightarrow 0}{\thicksim} -\frac{i}{\hbar} \left( E_\ell K_\ell + E_r K_r \right)
  \label{sc:firsttimeder}
\end{equation}
and
\begin{equation}
    \ddot{\rho} \underset{\hbar\rightarrow 0}{\thicksim} \frac{1}{\hbar^2} \left( \hbar \pder{E_\ell}{\tau_\ell}K_\ell + \hbar\pder{E_r}{\tau_r}K_r - E_\ell^2 K_\ell -E_r^2 K_r \right).
  \label{sc:secondtimeder}
\end{equation}
The combination of these two equations and Eq.~\eqref{1d:cddddot} gives the energy variance of $C_\mathrm{dd}$ as 
\begin{equation}
  \begin{split}
        \Delta H_\mathrm{dd}^2 \underset{\hbar\rightarrow 0}{\thicksim} &\frac{K_\ell K_r\left(E_\ell - E_r\right)^2}{(K_\ell + K_r)^2}\\
        &-\frac{\hbar}{K_\ell + K_r}\left(\pder{E_\ell}{\tau_\ell}K_\ell + \pder{E_r}{\tau_r}K_r \right).
  \end{split}
  \label{sc:semiclassicalhdd}
\end{equation}
A similar analysis gives $\Delta H_\mathrm{ff}$ in terms of the semiclassical quantities:
\begin{equation}
    \Delta H_\mathrm{ff}^2 \underset{\hbar\rightarrow 0}{\thicksim}
    \frac{1}{2} \left[ (E_\ell - E_r)^2 - \hbar \left(\pder{E_\ell}{\tau_\ell} + \pder{E_r}{\tau_r}\right) \right].
  \label{sc:semiclassicalhff}
\end{equation}
The final expressions for the semiclassical limits of the QI and 2OCE rates are then simply obtained by inserting either Eqs.~\eqref{sc:semiclassicalcff} and \eqref{sc:semiclassicalhdd} into Eq.~\eqref{1d:quantuminstanton} for the QI result, or Eqs.~\eqref{sc:semiclassicalcff} and \eqref{sc:semiclassicalhff} into Eq.~\eqref{1d:2oce} for the 2OCE result.

\begin{figure}[t]
	\includegraphics{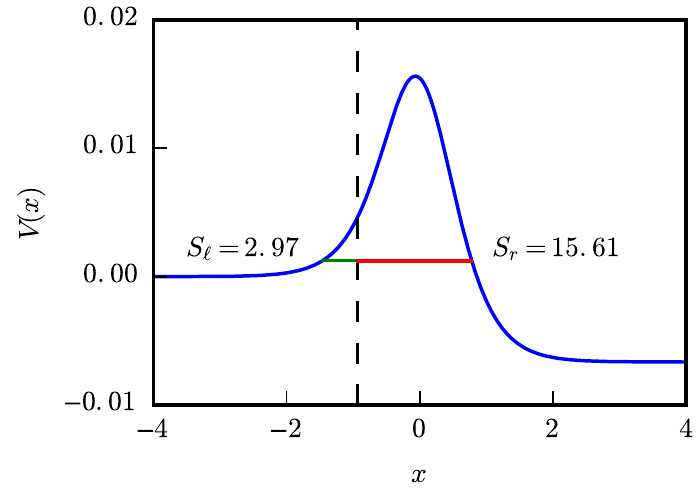}
	\caption{Two semiclassical minimum-action trajectories at 150 K, with $\alpha=1.425$, and with dividing surfaces chosen at $x_1=x_2=-0.930$ such that two paths combine to form an instanton.}
	\label{fig:energy_matching}
\end{figure}

Note that for less
asymmetric systems or at higher temperatures (above 142 K in this case), it is possible to find a new definition for the dividing surface for which the two minimum-action paths have the same energy and thus describe the instanton orbit, as shown in Fig.~\ref{fig:energy_matching}. However, despite the fact that these two paths together describe the correct periodic orbit,
we find that the QI approximation nonetheless overpredicts the rate by many orders of magnitude. This is because the left-bouncing path has a much smaller action than the right-bouncing path. As a result, the semiclassical analysis of $C_\mathrm{ff}(0)$ given in Eq.~\eqref{sc:semiclassicalcff} is no longer a good estimate of the quantum-mechanical value. The breakdown of the semiclassical analysis is due to the fact that
the term proportional to $K_\ell^2$ only 
cancels to first order in $\hbar$ and cannot be neglected when compared with the much smaller $K_\ell K_r$ term.
This problem is avoided by choosing dividing surfaces according to the standard prescriptions in Table~\ref{tab:surfaces}, for which $K_\ell$ and $K_r$ are of a similar order of magnitude, for which the semiclassical analysis is valid.

For a completely symmetric system at any temperature it is always possible to split the instanton in the middle, thereby generating two minimum-action paths of equal energy with $K_\ell=K_r$.
In the following, we shall use this fact to explain the success of the QI method for symmetric systems.

\subsection{Connection to the semiclassical instanton theory} \label{sec:connection}
The semiclassical limits to the QI and 2OCE approximations derived above show that the predicted rates do not just depend on the instanton path, but instead have contributions from the spurious path, $X_r$. However, the SCI result for the rate depends only on the instanton periodic orbit.\cite{InstReview} The implication is that, contrary to the original conjecture, the SCI method is not a simple semiclassical approximation to the QI or 2OCE methods. In fact, it can be seen from both the above expressions and Fig.~\ref{fig:sc_paths} that the contributions from the unphysical right-hand path will dominate in many situations and lead to erroneous rates. The only exception, when using the standard dividing surfaces satisfying Eq.~\eqref{1d:divsurfaces}, is for perfectly symmetric barriers where the two paths combine into the instanton periodic orbit.

These problems would go away if $\rho$ were dominated by the semiclassical paths which together make up the instanton periodic orbit. This is of course already the case for a symmetric system. Here we will show that if this were generally the case then the SCI rate formula would be recovered as the asymptotic limit of the QI expression.

If the two paths together describe a periodic orbit, then we can use $E_\ell=E_r$ and hence $p^\ell_1=p^r_1=p_1$ and $p^\ell_2=p^r_2=p_2$. The flux-flux correlation function [Eq.~\eqref{sc:semiclassicalcff}] thus simplifies to
\begin{equation}
	C_\mathrm{ff}(0) \underset{\hbar\rightarrow 0}{\thicksim} \frac{2 p_1 p_2}{m^2}  K_\ell K_r
        \label{sc:instantoncff}
\end{equation}
and
\begin{equation}
    \Delta H_\mathrm{ff}^2 \underset{\hbar\rightarrow 0}{\thicksim} - \frac{\hbar}{2} \left( \pder{E_\ell}{\tau_\ell} +  \pder{E_r}{\tau_r} \right).
  \label{sc:deltaHff}
\end{equation}

In order to study the QI approximation based on $\Delta H_\mathrm{dd}$, we define dividing surfaces which obey Eq.~\eqref{1d:divsurfaces} in the semiclassical limit [see also  Eq.~\eqref{SCdrhodx}], which implies that
$K_\ell = K_r$. The semiclassical limit of the energy variance $\Delta H_\mathrm{dd}$ [Eq.~\eqref{sc:semiclassicalhdd}] becomes
identical to that of $\Delta H_\mathrm{ff}$ given in Eq.~\eqref{sc:deltaHff}.

Eqs. \eqref{sc:instantoncff} and \eqref{sc:deltaHff} can thus be used to obtain the semiclassical limit of the QI and 2OCE approximations. Using the known expressions (for the 1D case)\cite{Gutzwiller1990}
\begin{equation}
-\pders{S_\gamma}{x_1}{x_2} = \frac{ m^2}{ p_1 p_2} \left( \frac{\partial^2 W_\gamma}{\partial E_\gamma^2}\right)^{-1},
\label{momentumeikonal}
\end{equation}
and
\begin{equation}
\left(\pder[2]{W_\gamma}{E_\gamma}\right)^{-1} = -\pder{E_\gamma}{\tau_\gamma},
\end{equation}
Eq.~\eqref{sc:instantoncff} becomes
\begin{equation}
C_\mathrm{ff}(0) \underset{\hbar\rightarrow 0}{\thicksim} \frac{1}{\pi \hbar}
    \sqrt{ \pder{E_\ell}{\tau_\ell} \pder{E_r}{\tau_r} } e^{-(S_\ell+S_r)/\hbar}.
\label{sc:finalinstantoncff}
\end{equation}
Using\cite{Richardson2016}
\begin{equation}
    \hbar\left(\der{E}{\beta}\right)^{-1} = \left(\pder{E_\ell}{\tau_\ell}\right)^{-1} + \left(\pder{E_r}{\tau_r}\right)^{-1},
\label{energybeta}
\end{equation}
where $E$ is the energy of the instanton orbit,
we obtain 
\begin{equation}
k_\mathrm{sc} Q_\mathrm{r} = \frac{1}{\sqrt{2\pi \hbar^2}} \left(- \der{E}{\beta}\right)^{1/2} e^{-(S_\ell+S_r)/\hbar},
\label{finalsemiclassicalrate}
\end{equation}
which matches Miller's original SCI expression \cite{Miller1975} in the one-dimensional case, bearing in mind that the total action of the instanton periodic orbit is given by $S_\ell+S_r$. This equation is equal to the ImF instanton expression,\cite{Affleck1981}
which can be written 
in a number of equivalent ways\cite{Althorpe2011,InstReview} including the ring-polymer formulation.\cite{Richardson2009} Thus, if the right-hand and left-hand paths combine to form the instanton and the dividing surfaces are chosen to satisfy Eq.~\eqref{1d:divsurfaces}, the QI and 2OCE expressions reduce to the SCI rate in the semiclassical limit. In general, however, the two paths only join exactly into the instanton periodic orbit for a symmetric system, hence the QI method is a particularly accurate method for symmetric barriers. For asymmetric barriers, the QI rate cannot be directly connected to the SCI expression.

\section{Projected QI}
\label{section:projected}
In this section, we suggest a modification of the quantum instanton approach to avoid sampling the spurious paths, which we will call the Projected Quantum Instanton (PQI) method. This defines a novel method in the spirit of the quantum instanton but which should give accurate rate predictions even for strongly asymmetric systems in the low-temperature limit.

In order to fix the QI method for asymmetric systems, it will be necessary to require that the two minimum-action paths have matching energies and thus combine into the instanton periodic orbit. Yet, for very asymmetric systems, it is impossible to choose dividing surfaces such that both minimum-action paths of imaginary-time $\beta\hbar/2$ have the same energy. Therefore, we will need to relax the requirement that the left- and right-bouncing paths have equal imaginary-time lengths \cite{InstReview} and will therefore also need a projection scheme which can categorize any general path into left and right sets. This can be achieved by defining the projected propagator
\begin{equation}
    \hat{U}_\gamma(t-i\tau_\gamma) = e^{-i \hat{H} (t-i \tau_\gamma)/2\hbar} \hat{\mathcal{P}}_\gamma e^{-i \hat{H} (t-i\tau_\gamma)/2\hbar},
\end{equation}
where the projection operators are written in terms of the Heaviside step function, $\theta$, as
\begin{subequations}
\begin{align}
    \hat{\mathcal{P}}_\ell &= \theta(x_0-\hat{x}) = \int_{-\infty}^{x_0} d x' \ket{x'}\bra{x'} \\
    \hat{\mathcal{P}}_r &= \theta(\hat{x}-x_0) = \int_{x_0}^{\infty} dx' \ket{x'}\bra{x'}
    \end{align}
    \label{pqi:projectors}
\end{subequations}
such that
\begin{equation}
    \hat{U}_\ell(t-i\tau) + \hat{U}_r(t-i\tau) = e^{-i \hat{H} (t-i\tau)/\hbar}.
    \label{pqi:totalprojector}
\end{equation}
Paths projected in this way can thus be categorized depending on whether their central point is to the right or left of the point $x_0$.
The location of $x_0$ is somewhat arbitrary for the following arguments as long as it appears between the instanton turning points. This way, $\braket{x_2|\hat{U}_\ell(-i\tau_\ell)|x_1}$ will be dominated by the left-bouncing minimum-action path and $\braket{x_2|\hat{U}_r(-i\tau_r)|x_1}$ by the right-bouncing minimum-action path. The projection operator we have used is the simplest choice to implement in our simulations.
However,
there are many other possible definitions
which also pick out the correct semiclassical pathways. Note that the multidimensional extension of the approach follows directly by projecting along a reaction coordinate.

By inserting projection operators into the Miller--Schwartz--Tromp formula for the rate [Eq.~(\ref{1d:reactionrate})],
we will effectively
neglect contributions from pairs of paths which either both bounce to the left or to the right.
This is expected to be a good approximation because within a semiclassical analysis these paths 
would give a zero contribution to the rate.\cite{Richardson2016,InstReview}
We therefore propose to approximate the rate as
\begin{equation}
     k Q_\mathrm{r} \approx \half \int_{-\infty}^\infty dt \, C^\mathcal{P}_\mathrm{ff}(t),
     \label{pqi:kPQI}
\end{equation}
where
\begin{equation}
     C^\mathcal{P}_\mathrm{ff}(t) = 2 \,\Tr \!\left( \hat{U}_\ell(t-i\tau_\ell) \hat{F}_1 \hat{U}_r(-t-i\tau_r) \hat{F}_2\right),
     \label{pqi:cp}
\end{equation}
which is a general (non-symmetric and complex) correlation function whose form permits the imaginary time to be different for the two propagators, although their sum, $\tau_\ell + \tau_r = \beta\hbar$, is fixed. This extra flexibility allows us to choose an appropriate dividing surface for which both the left- and right-bouncing paths can be minimum-action paths and also join together to describe the instanton. This idea follows a similar approach to that used in a first-principles derivation of the SCI method.\cite{Richardson2016,InstReview} It is absolutely necessary to project onto left and right paths when using a non-symmetric split of the imaginary time to avoid finding the spurious minimum-action paths of a right-bouncing path in time $\tau_\ell$ and vice versa. Note that the factor of 2 in Eq.~\eqref{pqi:cp} accounts for the alternative ordering of the projection operators, which integrates to the same result. \cite{InstReview}

Equation~\eqref{pqi:kPQI} is an approximation in general but remarkably it reproduces the exact rate of a free particle
as shown in Appendix~\ref{section:freeparticle}. In this important limit, despite the fact that the projected and unprojected correlation functions are not equivalent, their integrals over time are identical.

The final expression for the PQI rate is the second-order cumulant expansion of
\begin{equation} \label{pqi:CPff}
\begin{split}
    C^\mathcal{P}_\mathrm{ff}(t) = \frac{\hbar^2}{2m^2} \left( \pders{\rho_\ell}{x_1}{x_2}\right. & \rho_r^\ast- \pder{\rho_\ell}{x_1}\pder{\rho_r^\ast}{x_2}\\
    - \pder{\rho_\ell}{x_2}&\pder{\rho_r^\ast}{x_1} + \left.\rho_\ell \pders{\rho_r^\ast}{x_1}{x_2} \right),
    \end{split}
\end{equation}
where the projected matrix elements are given by 
\begin{equation}
\rho_\gamma\equiv\rho_\gamma(x_1,x_2,t) = \braket{x_2|\hat{U}_\gamma(t-i\tau_\gamma)|x_1},
\label{pqi:projectedrho}
\end{equation}
and their time derivatives calculated using  $\dot{\rho}_{\gamma}=i \pder{\rho_{\gamma}}{\tau_\gamma}$ and $\ddot{\rho}_\gamma=-\frac{\partial^2\rho_\gamma}{\partial\tau_\gamma^2}$. The PQI rate expression is then defined identically to the 2OCE approximation [Eq.~\eqref{1d:2oce}] except that $C_\mathrm{ff}$ is replaced by $C_\mathrm{ff}^\mathcal{P}$ throughout.

By following a similar approach as in the previous section we obtain the semiclassical limit of $C^\mathcal{P}_\mathrm{ff}(0)$ equal to that of Eq.~\eqref{sc:instantoncff}, and of $\Delta H_\mathrm{ff}^2$ given by Eq.~\eqref{sc:deltaHff}. The only difference with the previous analysis is that in PQI the minimum-action paths really do describe the instanton periodic orbit and no extra assumptions need be made. One can therefore show that the semiclassical limit of the rate expression reduces to SCI using the results of Sec.~\ref{sec:connection}. This result is actually unsurprising, as the derivation of SCI follows similar lines of reasoning in Ref.~\onlinecite{InstReview}.

In principle one can use any choice of dividing surfaces for which the two halves of the instanton periodic orbit are both minimum-action paths. We chose to place both the dividing surfaces as well as $x_0$ at
the location of the barrier maximum, and found that this choice obeyed the rule in each case tested. One should then optimize $\tau_\ell$ (keeping $\tau_\ell+\tau_r=\beta\hbar$ fixed) until $\dot{C}^\mathcal{P}_\mathrm{ff}(0)=0$. However, the value of $\tau_\ell$ obtained directly from SCI was found to be an excellent approximation to this optimal value.

We note that the steps used to derive the PQI method share some similarity to Wolynes' nonadiabatic quantum-transition theory.\cite{Wolynes1987nonadiabatic}
In this approach, the two paths are forced to bounce left or right depending on a projection onto
the two electronic states and for asymmetric systems may have different imaginary times. The idea for deriving new rate theories by ensuring that the instanton is the dominant path contributing to the rate has also been used in previous work on QTST \cite{Mills1997QTST,Richardson2009,Hele2013} and most recently in Ref.~\onlinecite{GRQTST} for nonadiabatic rates.

The PQI method is also applicable to multidimensional problems and can be efficiently computed using path-integral Monte Carlo approach
with a simple extension to the standard methodology. The central bead of each path plays the role of the $x'$ variable in Eq.~\eqref{pqi:projectors} and, for example, only paths for which $x'<x_0$ should contribute to $\rho_\ell$. In this work, however, we implement the projection by integrating numerically over the allowed range of $x'$ and evaluate the imaginary-time propagator as described in Appendix~\ref{eckartappendix}.

\section{Results and Discussion}
\label{section:discussion}
In order to validate our semiclassical analysis of the quantum instanton and cumulant expansion, it is useful to compute the numerical values of the various terms making up these approximations and to compare the values obtained from quantum mechanics (using the eigenfunctions for the Eckart barrier discussed in Appendix \ref{eckartappendix}) and from the semiclassical approximation. The results of the quantum and semiclassical calculations for the same system and temperature as those used in Figs.~\ref{asymmetricrho12} and \ref{fig:sc_paths} are compared in Table~\ref{tab:quantities}. The semiclassical limit is seen to be very close to the quantum values for each of these choices of dividing surfaces, being at most a factor of two different. The discrepancy is of course larger for the dividing surface choice made in Fig.~\ref{fig:energy_matching}, which leads to $K_\ell\gg K_r$. The good agreement is also seen in Table~\ref{tab:rates}, where the QI and 2OCE rate predictions are compared using both quantum and semiclassical calculations. However, all these approximations are an order of magnitude larger than the exact rate constant, showing that the source of the error is in the QI or 2OCE rate formulae themselves. (The system and temperature were in fact specifically chosen to demonstrate this order of magnitude error.) This justifies our use of a semiclassical evaluation of the relevant quantities for analysing the QI results.
The PQI method predicts a rate of $4.34\times10^{-12}$ a.u.\ without optimization of the imaginary-time split (i.e.\  using the semiclassical ratio of $\tau_r/\tau_\ell\approx0.258$), but when it is optimized to $\tau_r/\tau_\ell\approx0.298$, the rate prediction is $4.85\times10^{-12}$ a.u., which is even closer to the exact result (c.f. caption Table III).

\begin{table*}[t]
  \caption{Numerical values (in a.u.) of various quantities used in the QI expressions
  are calculated using quantum-mechanical (QM) and semiclassical (SC) methods for the specified dividing surfaces. The parameters of the potential are the same as in Table \ref{tab:surfaces} and the temperature is 100 K.
  In each case, the semiclassical version is within a factor of two of the quantum result, confirming that our semiclassical analysis of the quantum instanton is valid.
  }
  \begin{ruledtabular}\begin{tabular}{lcccccccc}
      \multirow{2}{*}{Div.\ Surf.}&\multicolumn{2}{c}{$C_{\mathrm{dd}}(0)$} & \multicolumn{2}{c}{$C_{\mathrm{ff}}(0)$} &\multicolumn{2}{c}{$\Delta H_{\mathrm{dd}}$} & \multicolumn{2}{c}{$\Delta H_{\mathrm{ff}}$}\\
      \cline{2-3} \cline{4-5} \cline{6-7} \cline{8-9}
      & QM & SC & QM & SC& QM & SC & QM & SC\\
      \hline
      A & $5.40 \times 10^{-9}$ & $4.16 \times 10^{-9}$&$4.60 \times 10^{-14}$ & $3.88 \times 10^{-14}$& 0.00323 & 0.00323 & 0.00437 & 0.00448\\
      B & $5.12 \times 10^{-9}$ & $4.03 \times 10^{-9}$&$6.60 \times 10^{-14}$ & $5.59 \times 10^{-14}$& 0.00337 & 0.00334 & 0.00454 & 0.00460\\
      C & $1.44 \times 10^{-7}$ & $1.31 \times 10^{-7}$&$4.11 \times 10^{-14}$ & $3.07 \times 10^{-14}$& 0.00129 & 0.00106 & 0.00409 & 0.00441\\
      D & $1.58 \times 10^{-8}$ & $2.27 \times 10^{-8}$&$1.61 \times 10^{-14}$ & $3.00 \times 10^{-14}$& 0.00231 & 0.00295 & 0.00351 & 0.00403\\
  \end{tabular}\end{ruledtabular}
  \label{tab:quantities}
\end{table*}

\begin{figure}[t]
\centering
\includegraphics{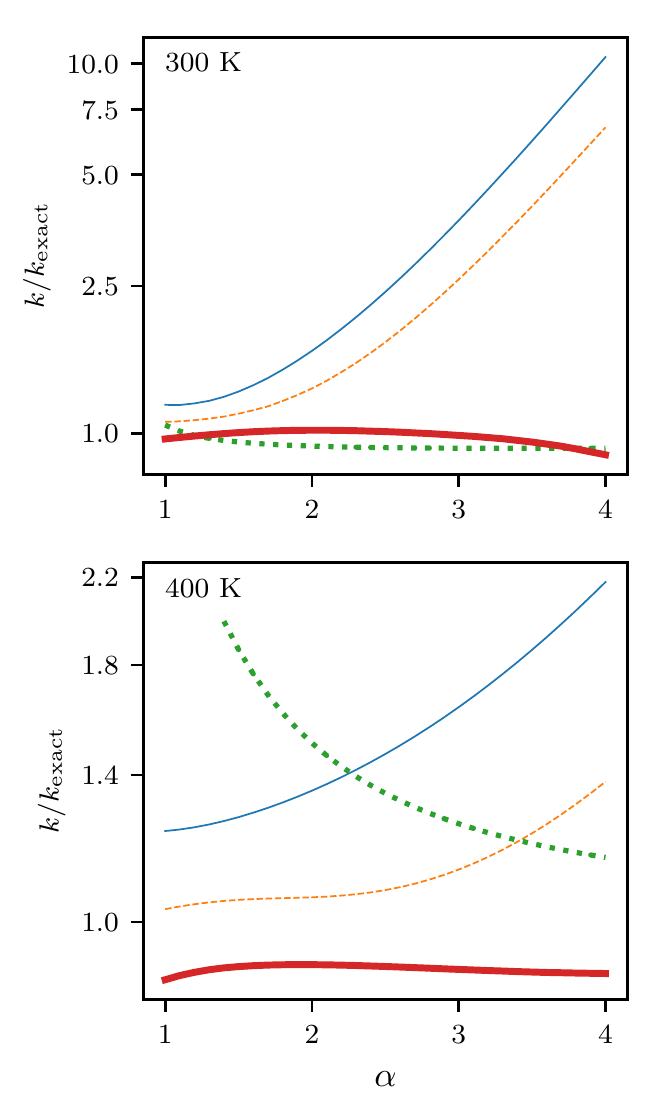}
\caption{\label{asymmetricerror}
Dependence of relative errors of various approximations for the rate constant on the asymmetry parameter $\alpha$ of the Eckart barrier for two different temperatures. The error is measured with the ratio $k/k_\mathrm{exact}$ (on a logarithmic scale for clarity), where $k_\mathrm{exact}$ and $k$ are, respectively, the exact and approximate rate constants. The different approximations shown are the quantum instanton with A-type dividing surfaces (thin blue line), the second-order cumulant expansion with A-type surfaces (thin orange dashed line), the semiclassical instanton (green dashed line), and the projected quantum instanton evaluated with dividing surfaces at the barrier maximum (thick red line). The SCI result is only plotted below the crossover temperature. 
}
\end{figure}

\begin{figure}[t]
\centering
\includegraphics{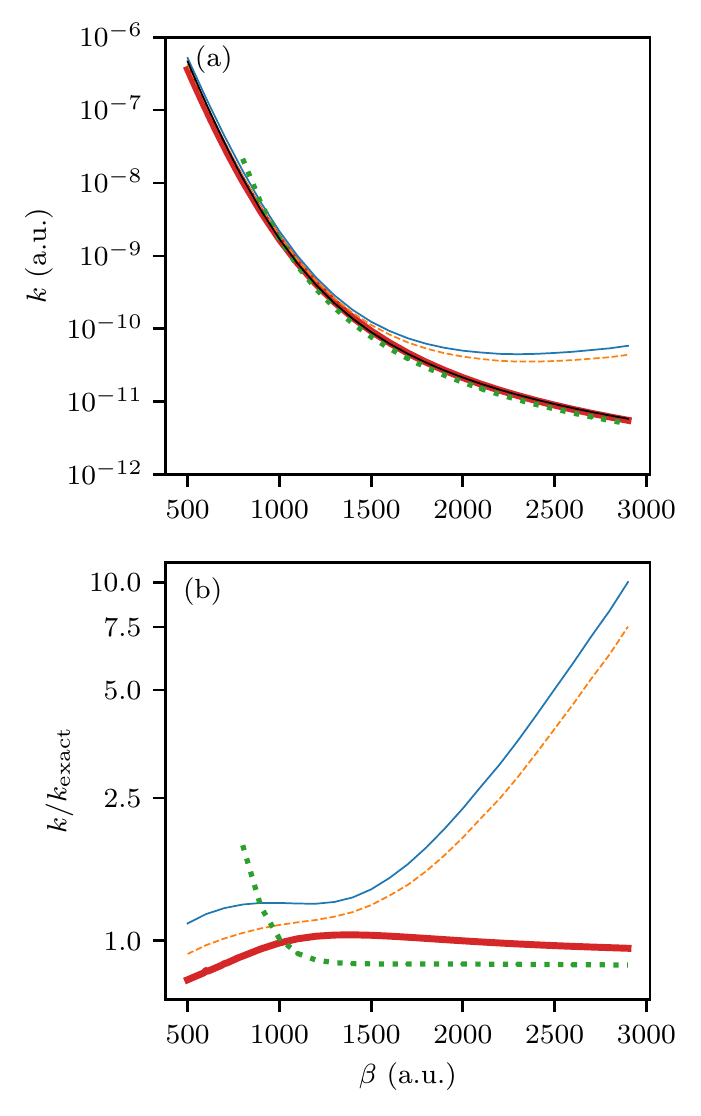}
\caption{\label{arrhenius} Arrhenius plots of the temperature dependence of (a) the rate constant and (b) the relative error of the rate constant evaluated with various approximations for a fixed asymmetry $\alpha=1.425$. The error and various line types are defined as in Fig.~\ref{asymmetricerror}. The crossover temperature for this system is 407 K ($\beta= 776$ a.u.).}
\end{figure}

The example system used in Fig.~\ref{fig:sc_paths} shows a significant error only at low enough temperatures.
However, we stress that the error seen in the QI method depends on both the asymmetry and the temperature, such that
for more asymmetric systems, even high temperatures can result in very large errors. For a more general outlook, Fig.~\ref{asymmetricerror} shows the relative error between the exact rate and the rates calculated using the four methods (QI, 2OCE, SCI, and PQI) as a function of asymmetry, $\alpha$, and for higher temperatures than that considered in Tables~\ref{tab:quantities} and \ref{tab:rates}.\footnote{The error in the QI rates is slightly larger than those reported in Ref.~\onlinecite{Miller2003} because they used an \emph{ad hoc} correction to the energy variance for their calculations, but this does not affect our conclusions.} To generate these results, we have generally favoured A-type dividing surfaces, although it would have been possible to choose surfaces for the 2OCE calculation that minimize the contributions of higher order terms.\cite{Ceotto2005} Nonetheless our semiclassical analysis implies that no dividing surface choice will significantly improve these results.

The temperature dependence of the exact and approximate rate constants is shown in Fig.~\ref{arrhenius}(a), along with the corresponding relative errors in Fig.~\ref{arrhenius}(b). While the SCI rates tend to a small constant relative error in the deep tunneling regime,\cite{Faraday, InstReview} the error in the QI rates increases exponentially with asymmetry. We attribute this growing error in the QI rate to the increasing contribution of the right-hand path in Fig.~\ref{fig:sc_paths}. In principle, 2OCE could be improved by going to a higher order expansion;\cite{Ceotto2005} however, in practice, very high orders will typically be required to cancel the spurious dominant contribution at zero time. In contrast, the PQI results in Figs.~\ref{asymmetricerror} and \ref{arrhenius} show a substantial improvement over both the QI and 2OCE results and are at least as accurate as SCI. For these PQI results, the value of $\tau_\ell$ is not optimized but simply chosen to match the imaginary time for the left bounce in the instanton trajectory.

Close to the cross-over temperature, $T_\mathrm{c} = \hbar \omega_\mathrm{b}/{2\pi k_\mathrm{B}}$ (where $\omega_\mathrm{b}$ is the absolute value of the imaginary barrier frequency),\cite{InstReview} the SCI approximation breaks down. The accuracy of the PQI rate is also decreased at higher temperatures, although 
it is not much worse than the original QI and 2OCE methods,
a statement which we can quantify by studying the rate of a free particle (and hence the high-temperature limit for arbitrary barriers).
In Appendix~\ref{section:freeparticle}, we calculate the PQI rate for this case and our analytical results explain why the PQI prediction slightly underestimates the exact rate in the high-temperature regime. However, we also show that direct integration of $C^\mathcal{P}_\mathrm{ff}(t)$ yields the exact rate for the free particle, suggesting that the only error lies in the steepest-descent approximation.

\begin{table}[t]
  \caption{Prediction of rates (in a.u.) using different methods, where $k_{\mathrm{2OCE}}$ is the rate from the cumulant expansion, and $k_{\mathrm{QI}}$ is the rate from the QI method. The parameters of the potential are the same as in Table \ref{tab:surfaces} and the temperature is 100 K.
  The exact rate is $4.63 \times 10^{-12}$ a.u.\ 
  and the semiclassical instanton rate is $3.93 \times 10^{-12}$ a.u.
  For comparison, the classical TST rate is $8.31 \times 10^{-26}$ a.u., which confirms that quantum tunneling effects are of significant importance. Rate constants were evaluated using either the quantum-mechanical (QM) or semiclassical (SC) components from Table~\ref{tab:quantities}.
}
 \begin{ruledtabular}
    \begin{tabular}{lcccc}
      \multirow{2}{*}{Div.\ Surf.} & \multicolumn{2}{c}{$k_{\mathrm{QI}} \times 10^{12}$} & \multicolumn{2}{c}{$k_{\mathrm{2OCE}}\times 10^{12}$}\\
      \cline{2-3} \cline{4-5}
      & QM & SC & QM & SC\\
      \hline
      A &54.6 &46.0 &40.3 &33.2 \\
      B & 75.1 &64.1 &55.7 &46.6 \\
      C & 123 &111 &38.6 &26.7 \\
      D & 26.7 &39.0  &17.5 &28.5 \\
    \end{tabular}
  \end{ruledtabular}
  \label{tab:rates}
\end{table}

Overall, these promising results show that the new PQI method is a substantial improvement over the traditional QI and 2OCE methods.
Future extensions of PQI for use with path-integral methods are therefore likely to fulfill the initial promise of QI as an extension to SCI, suitable for accurately predicting deep-tunneling reaction rates in highly anharmonic and asymmetric systems.

\section{Conclusions}
\label{section:conclusions}
In conclusion, we have presented a semiclassical analysis of the quantum instanton method. We have shown that the dominant contribution to the QI expression can arise from spurious paths, which leads to very large errors for very asymmetric barriers, especially at low temperatures. Consequently, despite conventional wisdom suggesting otherwise, no choice of dividing surface can remove this problem.
We justify our analysis by showing that semiclassical evaluation of the relevant quantities yields very similar numerical values to those from exact quantum mechanics for the asymmetric Eckart barrier. The major discrepancy from the exact rate therefore lies with the QI and 2OCE approximations themselves.

From our analysis we conclude that the SCI rate is the semiclassical limit of the QI rate \emph{only} when the barrier is perfectly symmetric. Therefore, the QI method provides an accurate prediction of the quantum rate for symmetric barriers. However, we find that SCI is much more accurate than QI at describing tunneling in asymmetric systems, although the SCI clearly cannot describe anharmonicity as well as the QI does.

Our proposed new method, PQI, fixes the inherent path-sampling problems with the QI and 2OCE methods mentioned above, and, at the same time, takes into account anharmonicities that SCI neglects. There exists some similarity between the PQI and theories employed to simulate electron-transfer reactions \cite{Wolynes1987nonadiabatic,GRQTST} and we hope that this study will inspire future development of nonadiabatic rate theories. It remains to be seen whether a path-integral application of PQI will become a competitive method for reaction rate calculations of complex systems.
Although the theory can be easily generalized to multidimensional systems,
it may not be so easy to apply to atomistic simulations of reactions in solution where SCI theory is not well defined.\cite{Perspective}
RPMD rate theory\cite{Craig2005b} also dominantly samples instanton paths even for asymmetric systems\cite{Richardson2009} and, in addition, includes classical recrossing effects neglected by SCI, QI, and PQI. It thus remains the method of choice for atomistic simulations
of reactions in solution.\cite{Habershon2013RPMDreview}

\section{Acknowledgements}
The authors acknowledge support from the Swiss National Science Foundation through the NCCR MUST (Molecular Ultrafast Science and Technology) Network and Dr. Konstantin Karandashev and Joseph Lawrence for useful discussions. M.J.T. is supported by an ETH Zurich Research Grant.

\appendix
\section{PQI, 2OCE, and QI rates for the free particle}
\label{section:freeparticle}
It is informative to examine the behaviour of PQI for the case of the free particle, which can be treated analytically.
The exact propagator for the free particle is well known\cite{Feynman1965}
\begin{equation}
\langle x | e^{-i H t/\hbar} | x' \rangle = \sqrt{\frac{m}{2\pi i \hbar t}} \exp\left[ \frac{i m (x - x')^2}{2 \hbar t}\right].
\label{pqi:freeparticlepropagator}
\end{equation}
If we substitute $t \rightarrow -i \tau_\gamma$ in Eq.~\eqref{pqi:freeparticlepropagator} and then insert Eq.~\eqref{pqi:freeparticlepropagator} into Eq.~\eqref{pqi:projectedrho}, we find
\begin{subequations}
\begin{align}
\rho_\ell &= \frac{m}{\pi \hbar \tau_\ell} \exp\left[-\frac{m}{\hbar \tau_\ell}(x_1^2 + x_2^2)\right] \nonumber\\
&\quad\times\int_{-\infty}^{x_0}\! dx' \,\exp\left\lbrace -\frac{2m}{\hbar \tau_\ell}\left[ (x')^2 - x'(x_1 + x_2)\right]\right\rbrace\\
\rho_r &= \frac{m}{\pi \hbar \tau_r} \exp\left[-\frac{m}{\hbar \tau_r}(x_1^2 + x_2^2)\right] \nonumber\\
&\quad\times\int^{\infty}_{x_0} \!dx' \,\exp\left\lbrace -\frac{2m}{\hbar \tau_r}\left[ (x')^2 - x'(x_1 + x_2)\right]\right\rbrace.
\end{align}
\label{pqi:freeparticlerho}
\end{subequations}
The integrals over $x'$ can be done analytically with standard results\cite{Gradshteyn} to obtain
\begin{subequations}
\begin{align}
\rho_\ell &= \frac{\rho }{2}\left[ 1 + \Phi(z_\ell) \right]\\
\rho_r &=  \frac{\rho }{2}\left[ 1 - \Phi(z_r) \right],
\end{align}
\label{pqi:analyticrho}
\end{subequations}
where
$\Phi(z)$ is the Gauss error function, and
\begin{equation}
z_\gamma = \sqrt{\frac{m}{2\hbar \tau_\gamma}}(2x_0-x_1 - x_2).
\end{equation}
Note that, as required from Eq.~\eqref{pqi:totalprojector}, $\rho = \rho_\ell + \rho_r$.

The free-particle delta-delta correlation function is
\begin{equation}
C_\mathrm{dd}(t)=\frac{m}{\pi \hbar \sqrt{4t^2 + \hbar^2 \beta^2}}\exp\left[\frac{-2m \beta (x_1-x_2)^2}{4t^2 + \hbar^2 \beta^2}\right]
\label{freeparticleCdd}
\end{equation}
and the stationarity condition [Eq.~(\ref{1d:divsurfaces})] for the dividing surfaces, equivalent to setting $\partial C_\mathrm{dd}(0)/ \partial x_{\gamma}=0 $, implies $x_1=x_2$. Taking $x_0=x_1=x_2$, we find $\rho_\ell  = \rho_r =\rho/2$.

For $x_1=x_2$, the exact quantum flux-flux correlation function for the free particle is \cite{Miller1983}
\begin{equation}
\label{freeparticleCff}
C_\mathrm{ff}(t)=\frac{1}{2\pi \hbar \beta}\frac{2\hbar^2 \beta^2}{\big[4t^2+\hbar^2 \beta^2\big]^{3/2}}.
\end{equation}
Quantities required for the QI rate [Eq.~(\ref{1d:quantuminstantonsp})] are thus
$\Delta H_\mathrm{dd} = \sqrt{2}/\beta$
and
$C_\mathrm{ff}(0)= (\pi \hbar^2 \beta^2)^{-1}$.
The resulting QI rate,
\begin{equation}
k_\text{QI} Q_\mathrm{r} = \frac{1}{2 \pi \hbar \beta} \sqrt{\frac{\pi}{2}},
\end{equation}
is a factor of $\sqrt{\pi/2}$ from the correct free-particle result,
$kQ_\mathrm{r}= (2 \pi \hbar \beta)^{-1}$, leading to an overestimation of the rate by 25$\%$.\cite{Miller2003QI}

For the 2OCE rate (\ref{1d:2oce}), the relevant quantity is instead
$\Delta H_\mathrm{ff} = \sqrt{6}/\beta$,
giving the rate
\begin{equation}
k_\text{2OCE} Q_\mathrm{r} = \frac{1}{2 \pi \hbar \beta} \sqrt{\frac{\pi}{6}},
\end{equation}
which differs from the exact free particle rate by a factor $\sqrt{\pi/6}$ (an underestimation by about 28$\%$).

Likewise, we can calculate the projected correlation function for the free particle using Eqs.~\eqref{pqi:CPff} and \eqref{pqi:analyticrho} with
the substitution $\tau_{\gamma} \rightarrow \tau_{\gamma} + i t$ to obtain
\begin{equation}
\label{freepartansatz}
C^\mathcal{P}_\mathrm{ff}(t)=\frac{1}{2 \pi \hbar \beta}\frac{\hbar^2 \beta^2}{\big[4t^2+\hbar^2 \beta^2\big]^{3/2}}+\frac{1}{\pi^2\big(4t^2+\hbar^2 \beta^2\big)}.
\end{equation}
The relevant quantities needed to calculate the PQI rate for the free particle are therefore

\begin{equation}
C^\mathcal{P}_\mathrm{ff}(0)= \frac{2+\pi}{2  \pi^2 \hbar^2 \beta^2}
\end{equation}
and
\begin{equation}
\Delta H^\mathcal{P}_\mathrm{ff} = \frac{\sqrt{2}}{\beta}\sqrt{\frac{4+3 \pi}{2+\pi}}.
\end{equation}
Therefore, the PQI rate for the free particle is given as 
\begin{equation}
k_\text{PQI} Q_\mathrm{r} = \frac{1}{2 \pi \hbar \beta} \frac{\sqrt{(2+\pi)^3}}{\sqrt{8 \pi (4+3 \pi)}},
\end{equation}
which underestimates the exact rate by about 37\%.

It is possible to directly integrate $ C^\mathcal{P}_\mathrm{ff}(t)$ over $t$ to yield the exact rate for this system. 
Then using Eq.~\eqref{pqi:kPQI}, we get $k Q_\mathrm{r} = (2 \pi \hbar \beta)^{-1}$, which is the exact rate for this system. \cite{Miller2003}
Therefore, for the special case of the free particle, no approximation is made by adding projections in Eq.~\eqref{pqi:kPQI}.
However, in their simplest forms, QI, 2OCE and PQI each give a different result for the rate of a free particle, none of which is correct. 
The errors in these approximate methods stem from the fact that neither the standard nor the projected flux-flux correlation function is a Gaussian function of $t$, and thus the steepest-descent approximation introduces an error.
This is well understood from the original QI work and
simple corrections have been suggested \cite{Miller2003,Ceotto2005,Hansen1996QTST} which could also be applied to improve the PQI rate formula.

\section{Eckart Barrier Eigenfunctions}
\label{eckartappendix}
In this appendix we derive closed-form expressions for the wavefunctions of the one-dimensional Eckart barrier. Eckart\cite{Eckart1930}
presented a derivation for only one wavefunction, which was enough to obtain the reflection and transmission coefficients. For our purpose, however, we will need two orthogonal wavefunctions to serve as a complete basis. In nearly all other respects, we follow his approach.

The Schr\"{o}dinger equation for this problem can be written as
\begin{equation}
	\xi^2 \pder[2]{u}{\xi} + \xi \pder{u}{\xi} + \frac{ml^2}{2 \pi^2 \hbar^2} \left[ \frac{A\xi}{1-\xi} + \frac{B\xi}{(1-\xi)^2} + E \right] u = 0
\end{equation}
where comparison with Eq.~\eqref{asymmetriceckart} gives $\xi=-e^{2\pi x/l}$, $A=V_0(1-\alpha)$, $B=V_0 (1 + \sqrt{\alpha})^2$ and $l=\pi/a$. Following Eckart, we define
\begin{subequations}
\begin{align}
\bar{\alpha} &= \sqrt{\frac{Eml^2}{2 \pi^2 \hbar^2}} \label{sqrtE} \\
\bar{\beta} &= \sqrt{\frac{(E-A)ml^2}{2 \pi^2 \hbar^2}}\\
\bar{\delta} &= \sqrt{\frac{2Bml^2 - \pi^2 \hbar^2}{4\pi^2 \hbar^2}}
\end{align}
\end{subequations}
and then rearrange into the form
\begin{equation}
\begin{split}
	&\xi^2(1-\xi)^2 \pder[2]{u}{\xi} + \xi(1-\xi)^2 \pder{u}{\xi} \\
	&+ \left[ \bar{\alpha}^2 + (-\bar{\beta}^2-\bar{\alpha}^2+\bar{\delta}^2+1/4)\xi + \bar{\beta}^2\xi^2\right] u = 0.
	\end{split}
	\label{appendix:schrodinger}
\end{equation}
Note that if $E<0$, the root in Eq.~\eqref{sqrtE} is taken such that $\Im\bar{\alpha}<0$. We assume that $A<0$, i.e. the asymptote on the right hand side is lower than that on the left.

It is known that the hypergeometric function solves Eq.~\eqref{appendix:schrodinger}. Klein \cite{Klein} showed that solutions to this differential equation can be obtained using the ansatz 
\begin{equation}
u(\xi) = \xi^{-i\bar{\alpha}} (1-\xi)^{1/2+i\bar{\delta}} F[a_0,b_0,c_0,\xi],
\end{equation}
with
\begin{subequations}
\begin{align}
a_0&=\thalf+i(-\bar{\alpha}+\bar{\beta}+\bar{\delta})\\
b_0&=\thalf-i(\bar{\alpha}+\bar{\beta}-\bar{\delta})\\
c_0&=1-2i\bar{\alpha}.
\end{align}
\end{subequations}
However, $F[a_0,b_0,c_0,\xi]$ is only one possible solution to the hypergeometric differential equation, and instead we choose
to use Forsyth's solutions \cite{Forsyth} called III and XIII to give 
\begin{widetext}
\begin{subequations}
\begin{align}
	u_1(\xi) &= (-1)^{-i\bar{\alpha}} \xi^{i\bar{\alpha}} (1-\xi)^{1/2+i\bar{\delta}} F[\thalf+i(\bar{\alpha}-\bar{\beta}+\bar{\delta}), \thalf+i(\bar{\alpha}+\bar{\beta}+\bar{\delta}), 1+2i\bar{\alpha}, \xi]\\
	u_2(\xi) &= (-1)^{i\bar{\alpha}} \xi^{-i\bar{\alpha}} (1-\xi)^{i(\bar{\alpha}-\bar{\beta})} F[\thalf+i(-\bar{\alpha}+\bar{\beta}+\bar{\delta}), -i/2(i+2\bar{\alpha}-2\bar{\beta}+2\bar{\delta}), 1+2i\bar{\beta}, 1/(1-\xi)].
\end{align}
\end{subequations}
\end{widetext}
The former is only a solution if $E>A$ and the second if $E>0$.
In order to find the $\xi\rightarrow-\infty$ limit, we first apply Abramowitz and Stegun's formula 15.3.7,\cite{Abramowitz} and after premultiplying by a constant we find the desired asymptotic limits for $u_1$ to be
\begin{subequations}
\begin{align}
	& \lim_{\xi\rightarrow 0} u_1(\xi) = e^{+i k_{\bar{\alpha}} x}
	\\
	& \lim_{\xi\rightarrow-\infty} u_1(\xi) = C_1 e^{-i k_{\bar{\beta} x}} + D_1 e^{+i k_{\bar{\beta}} x},
\end{align}
\end{subequations}
where
\begin{subequations}
\begin{align}
	C_1 &= \frac{\Gamma(1+2i\bar{\alpha})\Gamma(-2i\bar{\beta})}{\Gamma(\thalf+i(\bar{\alpha}-\bar{\beta}-\bar{\delta}))\Gamma(\thalf+i(\bar{\alpha}-\bar{\beta}+\bar{\delta}))}
	\\
	D_1 &= \frac{\Gamma(1+2i\bar{\alpha})\Gamma(2i\bar{\beta})}{\Gamma(\thalf+i(\bar{\alpha}+\bar{\beta}-\bar{\delta}))\Gamma(\thalf+i(\bar{\alpha}+\bar{\beta}+\bar{\delta}))}
\end{align}
\end{subequations}
and $k_{\bar{\alpha}}=2\pi\bar{\alpha}/l$ and $k_{\bar{\beta}}=2\pi\bar{\beta}/l$.
The asymptotic limits of the second wavefunction are (using Abramowitz and Stegun 15.3.9)\cite{Abramowitz}
\begin{subequations}
\begin{align}
	& \lim_{\xi\rightarrow-\infty} u_2(\xi) = e^{-i k_{\bar{\beta}} x}
	\\
	& \lim_{\xi\rightarrow 0} u_2(\xi) = A_2e^{-i k_{\bar{\alpha} x}} + B_2e^{+i k_{\bar{\alpha}} x},
\end{align}
\end{subequations}
where
\begin{subequations}
\begin{align}
	A_2 &= \frac{\Gamma(2i\bar{\alpha})\Gamma(1+2i\bar{\beta})}{\Gamma(\thalf+i(\bar{\alpha}+\bar{\beta}-\bar{\delta}))\Gamma(\thalf+i(\bar{\alpha}+\bar{\beta}+\bar{\delta}))}
	\\
	B_2 &= \frac{\Gamma(-2i\bar{\alpha})\Gamma(1+2i\bar{\beta})}{\Gamma(\thalf+i(-\bar{\alpha}+\bar{\beta}-\bar{\delta}))\Gamma(\thalf+i(-\bar{\alpha}+\bar{\beta}+\bar{\delta}))}.
\end{align}
\end{subequations}
Note that Eckart's results for reflection and transmission are given by $|B_2/A_2|^2$ and $|\bar{\beta}/\bar{\alpha}|/|A_2|^2$ given a particle incident on the left.

Finally, we normalize the wavefunctions
\begin{subequations}
\begin{align}
\psi_1(x;E) &= \frac{u_1(\xi(x))}{\sqrt{\pi N_1}}\\
\psi_2(x;E) &= \frac{u_2(\xi(x))}{\sqrt{\pi N_2}},
\end{align}
\end{subequations}
such that 
\begin{equation}
\int\! dx \, \psi_1(x;E)^* \psi_1(x;E') = \delta(E-E').
\end{equation}
Using 
\begin{equation}
\frac{1}{\pi}\int_0^\infty \! dx\, e^{i [f(E)-f(E')] x}= \left|\pder{f}{E}\right|^{-1} \delta(E-E'),
\end{equation}
we find the normalization constants should be chosen as
\begin{subequations}
\begin{align}
	N_1 &= \frac{1}{k_{\bar{\alpha}}'} + \frac{|C_1|^2}{k_{\bar{\beta}}'} + \frac{|D_1|^2}{k_{\bar{\beta}}'}
	\\
	N_2 &= \frac{1}{k_{\bar{\beta}}'} + \frac{|A_2|^2}{k_{\bar{\alpha}}'} + \frac{|B_2|^2}{k_{\bar{\alpha}}'},
\end{align}
\end{subequations}
where $k_{\bar{\alpha}}'=\pder{k_{\bar{\alpha}}}{E}=k_{\bar{\alpha}}/(2E)$ and $k_{\bar{\beta}}'=\pder{k_{\bar{\beta}}}{E}=k_{\bar{\beta}}/(2(E-A))$.
The two wavefunctions must be orthogonal because one corresponds to an incoming particle from the left and the other from the right.
Mathematically, it occurs because $B_2/k_{\bar{\alpha}}' + C_1^*/k_{\bar{\beta}}'=0$.

The propagator can be evaluated numerically as an integral over energies using
\begin{equation}
\begin{split}
\braket{x_2|e^{-i \hat{H}t/\hbar}|x_1} &=
\int_A^\infty \! dE\, \psi_1(x_2;E)\psi_1(x_1;E)^* e^{-i E t/\hbar} \\
&+\int_0^\infty \! dE \, \psi_2(x_2;E)\psi_2(x_1;E)^* e^{-i E t/\hbar},
\end{split}
\end{equation}%
where $t$ can also be complex in order to obtain the density matrix. Derivatives of the propagator can be evaluated by explicit differentiation of the wavefunctions, although 
in practice we found that it was simpler and sufficiently accurate to evaluate them with finite differences.

\bibliography{quantuminstanton,jeremy,references}

\begin{thebibliography}{64}%
\makeatletter
\providecommand \@ifxundefined [1]{%
 \@ifx{#1\undefined}
}%
\providecommand \@ifnum [1]{%
 \ifnum #1\expandafter \@firstoftwo
 \else \expandafter \@secondoftwo
 \fi
}%
\providecommand \@ifx [1]{%
 \ifx #1\expandafter \@firstoftwo
 \else \expandafter \@secondoftwo
 \fi
}%
\providecommand \natexlab [1]{#1}%
\providecommand \enquote  [1]{``#1''}%
\providecommand \bibnamefont  [1]{#1}%
\providecommand \bibfnamefont [1]{#1}%
\providecommand \citenamefont [1]{#1}%
\providecommand \href@noop [0]{\@secondoftwo}%
\providecommand \href [0]{\begingroup \@sanitize@url \@href}%
\providecommand \@href[1]{\@@startlink{#1}\@@href}%
\providecommand \@@href[1]{\endgroup#1\@@endlink}%
\providecommand \@sanitize@url [0]{\catcode `\\12\catcode `\$12\catcode
  `\&12\catcode `\#12\catcode `\^12\catcode `\_12\catcode `\%12\relax}%
\providecommand \@@startlink[1]{}%
\providecommand \@@endlink[0]{}%
\providecommand \url  [0]{\begingroup\@sanitize@url \@url }%
\providecommand \@url [1]{\endgroup\@href {#1}{\urlprefix }}%
\providecommand \urlprefix  [0]{URL }%
\providecommand \Eprint [0]{\href }%
\providecommand \doibase [0]{http://dx.doi.org/}%
\providecommand \selectlanguage [0]{\@gobble}%
\providecommand \bibinfo  [0]{\@secondoftwo}%
\providecommand \bibfield  [0]{\@secondoftwo}%
\providecommand \translation [1]{[#1]}%
\providecommand \BibitemOpen [0]{}%
\providecommand \bibitemStop [0]{}%
\providecommand \bibitemNoStop [0]{.\EOS\space}%
\providecommand \EOS [0]{\spacefactor3000\relax}%
\providecommand \BibitemShut  [1]{\csname bibitem#1\endcsname}%
\let\auto@bib@innerbib\@empty
\bibitem [{\citenamefont {Miller}, \citenamefont {Schwartz},\ and\
  \citenamefont {Tromp}(1983)}]{Miller1983}%
  \BibitemOpen
  \bibfield  {author} {\bibinfo {author} {\bibfnamefont {W.~H.}\ \bibnamefont
  {Miller}}, \bibinfo {author} {\bibfnamefont {S.~D.}\ \bibnamefont
  {Schwartz}}, \ and\ \bibinfo {author} {\bibfnamefont {J.~W.}\ \bibnamefont
  {Tromp}},\ }\href@noop {} {\bibfield  {journal} {\bibinfo  {journal} {J.
  Chem. Phys.}\ }\textbf {\bibinfo {volume} {79}},\ \bibinfo {pages} {4889}
  (\bibinfo {year} {1983})}\BibitemShut {NoStop}%
\bibitem [{\citenamefont {Miller}(2001{\natexlab{a}})}]{Miller2001}%
  \BibitemOpen
  \bibfield  {author} {\bibinfo {author} {\bibfnamefont {W.~H.}\ \bibnamefont
  {Miller}},\ }\href@noop {} {\bibfield  {journal} {\bibinfo  {journal} {J.
  Phys. Chem. A}\ }\textbf {\bibinfo {volume} {105}},\ \bibinfo {pages} {2942}
  (\bibinfo {year} {2001}{\natexlab{a}})}\BibitemShut {NoStop}%
\bibitem [{\citenamefont {Cao}\ and\ \citenamefont {Voth}(1994)}]{Cao1994}%
  \BibitemOpen
  \bibfield  {author} {\bibinfo {author} {\bibfnamefont {J.}~\bibnamefont
  {Cao}}\ and\ \bibinfo {author} {\bibfnamefont {G.~A.}\ \bibnamefont {Voth}},\
  }\href@noop {} {\bibfield  {journal} {\bibinfo  {journal} {J. Chem. Phys.}\
  }\textbf {\bibinfo {volume} {100}},\ \bibinfo {pages} {5106} (\bibinfo {year}
  {1994})}\BibitemShut {NoStop}%
\bibitem [{\citenamefont {Jang}\ and\ \citenamefont {Voth}(1999)}]{Jang1999}%
  \BibitemOpen
  \bibfield  {author} {\bibinfo {author} {\bibfnamefont {S.}~\bibnamefont
  {Jang}}\ and\ \bibinfo {author} {\bibfnamefont {G.~A.}\ \bibnamefont
  {Voth}},\ }\href@noop {} {\bibfield  {journal} {\bibinfo  {journal} {J. Chem.
  Phys.}\ }\textbf {\bibinfo {volume} {111}},\ \bibinfo {pages} {2371}
  (\bibinfo {year} {1999})}\BibitemShut {NoStop}%
\bibitem [{\citenamefont {Craig}\ and\ \citenamefont
  {Manolopoulos}(2004)}]{Craig2004}%
  \BibitemOpen
  \bibfield  {author} {\bibinfo {author} {\bibfnamefont {I.~R.}\ \bibnamefont
  {Craig}}\ and\ \bibinfo {author} {\bibfnamefont {D.~E.}\ \bibnamefont
  {Manolopoulos}},\ }\href@noop {} {\bibfield  {journal} {\bibinfo  {journal}
  {J. Chem. Phys.}\ }\textbf {\bibinfo {volume} {121}},\ \bibinfo {pages}
  {3368} (\bibinfo {year} {2004})}\BibitemShut {NoStop}%
\bibitem [{\citenamefont {Craig}\ and\ \citenamefont
  {Manolopoulos}(2005{\natexlab{a}})}]{Craig2005a}%
  \BibitemOpen
  \bibfield  {author} {\bibinfo {author} {\bibfnamefont {I.~R.}\ \bibnamefont
  {Craig}}\ and\ \bibinfo {author} {\bibfnamefont {D.~E.}\ \bibnamefont
  {Manolopoulos}},\ }\href@noop {} {\bibfield  {journal} {\bibinfo  {journal}
  {J. Chem. Phys.}\ }\textbf {\bibinfo {volume} {122}},\ \bibinfo {pages}
  {084106} (\bibinfo {year} {2005}{\natexlab{a}})}\BibitemShut {NoStop}%
\bibitem [{\citenamefont {Craig}\ and\ \citenamefont
  {Manolopoulos}(2005{\natexlab{b}})}]{Craig2005b}%
  \BibitemOpen
  \bibfield  {author} {\bibinfo {author} {\bibfnamefont {I.~R.}\ \bibnamefont
  {Craig}}\ and\ \bibinfo {author} {\bibfnamefont {D.~E.}\ \bibnamefont
  {Manolopoulos}},\ }\href@noop {} {\bibfield  {journal} {\bibinfo  {journal}
  {J. Chem. Phys.}\ }\textbf {\bibinfo {volume} {123}},\ \bibinfo {pages}
  {034102} (\bibinfo {year} {2005}{\natexlab{b}})}\BibitemShut {NoStop}%
\bibitem [{\citenamefont {Hele}\ \emph
  {et~al.}(2015{\natexlab{a}})\citenamefont {Hele}, \citenamefont {Willatt},
  \citenamefont {Muolo},\ and\ \citenamefont {Althorpe}}]{Hele2015a}%
  \BibitemOpen
  \bibfield  {author} {\bibinfo {author} {\bibfnamefont {T.~J.~H.}\
  \bibnamefont {Hele}}, \bibinfo {author} {\bibfnamefont {M.~J.}\ \bibnamefont
  {Willatt}}, \bibinfo {author} {\bibfnamefont {A.}~\bibnamefont {Muolo}}, \
  and\ \bibinfo {author} {\bibfnamefont {S.~C.}\ \bibnamefont {Althorpe}},\
  }\href@noop {} {\bibfield  {journal} {\bibinfo  {journal} {J. Chem. Phys.}\
  }\textbf {\bibinfo {volume} {142}},\ \bibinfo {pages} {134103} (\bibinfo
  {year} {2015}{\natexlab{a}})}\BibitemShut {NoStop}%
\bibitem [{\citenamefont {Hele}\ \emph
  {et~al.}(2015{\natexlab{b}})\citenamefont {Hele}, \citenamefont {Willatt},
  \citenamefont {Muolo},\ and\ \citenamefont {Althorpe}}]{Hele2015b}%
  \BibitemOpen
  \bibfield  {author} {\bibinfo {author} {\bibfnamefont {T.~J.~H.}\
  \bibnamefont {Hele}}, \bibinfo {author} {\bibfnamefont {M.~J.}\ \bibnamefont
  {Willatt}}, \bibinfo {author} {\bibfnamefont {A.}~\bibnamefont {Muolo}}, \
  and\ \bibinfo {author} {\bibfnamefont {S.~C.}\ \bibnamefont {Althorpe}},\
  }\href@noop {} {\bibfield  {journal} {\bibinfo  {journal} {J. Chem. Phys.}\
  }\textbf {\bibinfo {volume} {142}},\ \bibinfo {pages} {191101} (\bibinfo
  {year} {2015}{\natexlab{b}})}\BibitemShut {NoStop}%
\bibitem [{\citenamefont {Willatt}, \citenamefont {Ceriotti},\ and\
  \citenamefont {Althorpe}(2018)}]{Willatt2018}%
  \BibitemOpen
  \bibfield  {author} {\bibinfo {author} {\bibfnamefont {M.~J.}\ \bibnamefont
  {Willatt}}, \bibinfo {author} {\bibfnamefont {M.}~\bibnamefont {Ceriotti}}, \
  and\ \bibinfo {author} {\bibfnamefont {S.~C.}\ \bibnamefont {Althorpe}},\
  }\href@noop {} {\bibfield  {journal} {\bibinfo  {journal} {J. Chem. Phys.}\
  }\textbf {\bibinfo {volume} {148}},\ \bibinfo {pages} {102336} (\bibinfo
  {year} {2018})}\BibitemShut {NoStop}%
\bibitem [{\citenamefont {Trenins}\ and\ \citenamefont
  {Althorpe}(2018)}]{Trenins2018}%
  \BibitemOpen
  \bibfield  {author} {\bibinfo {author} {\bibfnamefont {G.}~\bibnamefont
  {Trenins}}\ and\ \bibinfo {author} {\bibfnamefont {S.~C.}\ \bibnamefont
  {Althorpe}},\ }\href@noop {} {\bibfield  {journal} {\bibinfo  {journal} {J.
  Chem. Phys.}\ }\textbf {\bibinfo {volume} {149}},\ \bibinfo {pages} {014102}
  (\bibinfo {year} {2018})}\BibitemShut {NoStop}%
\bibitem [{\citenamefont {Miller}(2001{\natexlab{b}})}]{Miller2001SCIVR}%
  \BibitemOpen
  \bibfield  {author} {\bibinfo {author} {\bibfnamefont {W.~H.}\ \bibnamefont
  {Miller}},\ }\href {\doibase 10.1021/jp003712k} {\bibfield  {journal}
  {\bibinfo  {journal} {J.~Phys. Chem.~A}\ }\textbf {\bibinfo {volume} {105}},\
  \bibinfo {pages} {2942} (\bibinfo {year} {2001}{\natexlab{b}})}\BibitemShut
  {NoStop}%
\bibitem [{\citenamefont {Miller}(1974)}]{Miller1974}%
  \BibitemOpen
  \bibfield  {author} {\bibinfo {author} {\bibfnamefont {W.~H.}\ \bibnamefont
  {Miller}},\ }\href@noop {} {\bibfield  {journal} {\bibinfo  {journal} {J.
  Chem. Phys.}\ }\textbf {\bibinfo {volume} {61}},\ \bibinfo {pages} {1823}
  (\bibinfo {year} {1974})}\BibitemShut {NoStop}%
\bibitem [{\citenamefont {Pollak}\ and\ \citenamefont
  {Liao}(1998)}]{Pollak1998}%
  \BibitemOpen
  \bibfield  {author} {\bibinfo {author} {\bibfnamefont {E.}~\bibnamefont
  {Pollak}}\ and\ \bibinfo {author} {\bibfnamefont {J.~L.}\ \bibnamefont
  {Liao}},\ }\href@noop {} {\bibfield  {journal} {\bibinfo  {journal} {J. Chem.
  Phys.}\ }\textbf {\bibinfo {volume} {108}},\ \bibinfo {pages} {2733}
  (\bibinfo {year} {1998})}\BibitemShut {NoStop}%
\bibitem [{\citenamefont {Miller}(1975)}]{Miller1975}%
  \BibitemOpen
  \bibfield  {author} {\bibinfo {author} {\bibfnamefont {W.~H.}\ \bibnamefont
  {Miller}},\ }\href@noop {} {\bibfield  {journal} {\bibinfo  {journal} {J.
  Chem. Phys.}\ }\textbf {\bibinfo {volume} {62}},\ \bibinfo {pages} {1899}
  (\bibinfo {year} {1975})}\BibitemShut {NoStop}%
\bibitem [{\citenamefont {Richardson}(2018{\natexlab{a}})}]{Perspective}%
  \BibitemOpen
  \bibfield  {author} {\bibinfo {author} {\bibfnamefont {J.~O.}\ \bibnamefont
  {Richardson}},\ }\href {\doibase 10.1063/1.5028352} {\bibfield  {journal}
  {\bibinfo  {journal} {J. Chem. Phys.}\ }\textbf {\bibinfo {volume} {148}},\
  \bibinfo {pages} {200901} (\bibinfo {year} {2018}{\natexlab{a}})}\BibitemShut
  {NoStop}%
\bibitem [{\citenamefont {Richardson}(2016{\natexlab{a}})}]{Richardson2016}%
  \BibitemOpen
  \bibfield  {author} {\bibinfo {author} {\bibfnamefont {J.~O.}\ \bibnamefont
  {Richardson}},\ }\href@noop {} {\bibfield  {journal} {\bibinfo  {journal} {J.
  Chem. Phys.}\ }\textbf {\bibinfo {volume} {144}},\ \bibinfo {pages} {114106}
  (\bibinfo {year} {2016}{\natexlab{a}})}\BibitemShut {NoStop}%
\bibitem [{\citenamefont {Richardson}(2018{\natexlab{b}})}]{InstReview}%
  \BibitemOpen
  \bibfield  {author} {\bibinfo {author} {\bibfnamefont {J.~O.}\ \bibnamefont
  {Richardson}},\ }\href {\doibase 10.1080/0144235X.2018.1472353} {\bibfield
  {journal} {\bibinfo  {journal} {Int. Rev. Phys. Chem.}\ }\textbf {\bibinfo
  {volume} {37}},\ \bibinfo {pages} {171} (\bibinfo {year}
  {2018}{\natexlab{b}})}\BibitemShut {NoStop}%
\bibitem [{\citenamefont {Affleck}(1981)}]{Affleck1981}%
  \BibitemOpen
  \bibfield  {author} {\bibinfo {author} {\bibfnamefont {I.}~\bibnamefont
  {Affleck}},\ }\href@noop {} {\bibfield  {journal} {\bibinfo  {journal} {Phys.
  Rev. Lett.}\ }\textbf {\bibinfo {volume} {46}},\ \bibinfo {pages} {388}
  (\bibinfo {year} {1981})}\BibitemShut {NoStop}%
\bibitem [{\citenamefont {Althorpe}(2011)}]{Althorpe2011}%
  \BibitemOpen
  \bibfield  {author} {\bibinfo {author} {\bibfnamefont {S.~C.}\ \bibnamefont
  {Althorpe}},\ }\href@noop {} {\bibfield  {journal} {\bibinfo  {journal} {J.
  Chem. Phys.}\ }\textbf {\bibinfo {volume} {134}},\ \bibinfo {pages} {114104}
  (\bibinfo {year} {2011})}\BibitemShut {NoStop}%
\bibitem [{\citenamefont {Andersson}\ \emph {et~al.}(2009)\citenamefont
  {Andersson}, \citenamefont {Nyman}, \citenamefont {Arnaldsson}, \citenamefont
  {Manthe},\ and\ \citenamefont {J{\'o}nsson}}]{Andersson2009Hmethane}%
  \BibitemOpen
  \bibfield  {author} {\bibinfo {author} {\bibfnamefont {S.}~\bibnamefont
  {Andersson}}, \bibinfo {author} {\bibfnamefont {G.}~\bibnamefont {Nyman}},
  \bibinfo {author} {\bibfnamefont {A.}~\bibnamefont {Arnaldsson}}, \bibinfo
  {author} {\bibfnamefont {U.}~\bibnamefont {Manthe}}, \ and\ \bibinfo {author}
  {\bibfnamefont {H.}~\bibnamefont {J{\'o}nsson}},\ }\href {\doibase
  10.1021/jp811070w} {\bibfield  {journal} {\bibinfo  {journal} {J.~Phys.
  Chem.~A}\ }\textbf {\bibinfo {volume} {113}},\ \bibinfo {pages} {4468}
  (\bibinfo {year} {2009})}\BibitemShut {NoStop}%
\bibitem [{\citenamefont {Richardson}\ and\ \citenamefont
  {Althorpe}(2009)}]{Richardson2009}%
  \BibitemOpen
  \bibfield  {author} {\bibinfo {author} {\bibfnamefont {J.~O.}\ \bibnamefont
  {Richardson}}\ and\ \bibinfo {author} {\bibfnamefont {S.~C.}\ \bibnamefont
  {Althorpe}},\ }\href@noop {} {\bibfield  {journal} {\bibinfo  {journal} {J.
  Chem. Phys.}\ }\textbf {\bibinfo {volume} {131}},\ \bibinfo {pages} {214106}
  (\bibinfo {year} {2009})}\BibitemShut {NoStop}%
\bibitem [{\citenamefont {Rommel}, \citenamefont {Goumans},\ and\ \citenamefont
  {K\"astner}(2011)}]{Rommel2011locating}%
  \BibitemOpen
  \bibfield  {author} {\bibinfo {author} {\bibfnamefont {J.~B.}\ \bibnamefont
  {Rommel}}, \bibinfo {author} {\bibfnamefont {T.~P.~M.}\ \bibnamefont
  {Goumans}}, \ and\ \bibinfo {author} {\bibfnamefont {J.}~\bibnamefont
  {K\"astner}},\ }\href {\doibase 10.1021/ct100658y} {\bibfield  {journal}
  {\bibinfo  {journal} {J.~Chem. Theory Comput.}\ }\textbf {\bibinfo {volume}
  {7}},\ \bibinfo {pages} {690} (\bibinfo {year} {2011})}\BibitemShut {NoStop}%
\bibitem [{\citenamefont {Vaillant}, \citenamefont {Althorpe},\ and\
  \citenamefont {Wales}(2019)}]{Vaillant2019}%
  \BibitemOpen
  \bibfield  {author} {\bibinfo {author} {\bibfnamefont {C.~L.}\ \bibnamefont
  {Vaillant}}, \bibinfo {author} {\bibfnamefont {S.~C.}\ \bibnamefont
  {Althorpe}}, \ and\ \bibinfo {author} {\bibfnamefont {D.~J.}\ \bibnamefont
  {Wales}},\ }\href@noop {} {\bibfield  {journal} {\bibinfo  {journal} {J.
  Chem. Theory Comput.}\ }\textbf {\bibinfo {volume} {15}},\ \bibinfo {pages}
  {33} (\bibinfo {year} {2019})}\BibitemShut {NoStop}%
\bibitem [{\citenamefont {Miller}\ \emph
  {et~al.}(2003{\natexlab{a}})\citenamefont {Miller}, \citenamefont {Zhao},
  \citenamefont {Ceotto},\ and\ \citenamefont {Yang}}]{Miller2003}%
  \BibitemOpen
  \bibfield  {author} {\bibinfo {author} {\bibfnamefont {W.~H.}\ \bibnamefont
  {Miller}}, \bibinfo {author} {\bibfnamefont {Y.}~\bibnamefont {Zhao}},
  \bibinfo {author} {\bibfnamefont {M.}~\bibnamefont {Ceotto}}, \ and\ \bibinfo
  {author} {\bibfnamefont {S.}~\bibnamefont {Yang}},\ }\href@noop {} {\bibfield
   {journal} {\bibinfo  {journal} {J. Chem. Phys.}\ }\textbf {\bibinfo {volume}
  {119}},\ \bibinfo {pages} {1329} (\bibinfo {year}
  {2003}{\natexlab{a}})}\BibitemShut {NoStop}%
\bibitem [{\citenamefont {Yamamoto}\ and\ \citenamefont
  {Miller}(2004)}]{Yamamoto2004}%
  \BibitemOpen
  \bibfield  {author} {\bibinfo {author} {\bibfnamefont {T.}~\bibnamefont
  {Yamamoto}}\ and\ \bibinfo {author} {\bibfnamefont {W.~H.}\ \bibnamefont
  {Miller}},\ }\href@noop {} {\bibfield  {journal} {\bibinfo  {journal} {J.
  Chem. Phys.}\ }\textbf {\bibinfo {volume} {120}},\ \bibinfo {pages} {3086}
  (\bibinfo {year} {2004})}\BibitemShut {NoStop}%
\bibitem [{\citenamefont {Zhao}, \citenamefont {Yamamoto},\ and\ \citenamefont
  {Miller}(2004)}]{Zhao2004}%
  \BibitemOpen
  \bibfield  {author} {\bibinfo {author} {\bibfnamefont {Y.}~\bibnamefont
  {Zhao}}, \bibinfo {author} {\bibfnamefont {T.}~\bibnamefont {Yamamoto}}, \
  and\ \bibinfo {author} {\bibfnamefont {W.~H.}\ \bibnamefont {Miller}},\
  }\href@noop {} {\bibfield  {journal} {\bibinfo  {journal} {J. Chem. Phys.}\
  }\textbf {\bibinfo {volume} {120}},\ \bibinfo {pages} {3100} (\bibinfo {year}
  {2004})}\BibitemShut {NoStop}%
\bibitem [{\citenamefont {Wang}\ and\ \citenamefont {Zhao}(2009)}]{Wang2009}%
  \BibitemOpen
  \bibfield  {author} {\bibinfo {author} {\bibfnamefont {W.}~\bibnamefont
  {Wang}}\ and\ \bibinfo {author} {\bibfnamefont {Y.}~\bibnamefont {Zhao}},\
  }\href@noop {} {\bibfield  {journal} {\bibinfo  {journal} {J. Chem. Phys.}\
  }\textbf {\bibinfo {volume} {130}},\ \bibinfo {pages} {114708} (\bibinfo
  {year} {2009})}\BibitemShut {NoStop}%
\bibitem [{\citenamefont {Yamamoto}\ and\ \citenamefont
  {Miller}(2005)}]{Yamamoto2005}%
  \BibitemOpen
  \bibfield  {author} {\bibinfo {author} {\bibfnamefont {T.}~\bibnamefont
  {Yamamoto}}\ and\ \bibinfo {author} {\bibfnamefont {W.~H.}\ \bibnamefont
  {Miller}},\ }\href@noop {} {\bibfield  {journal} {\bibinfo  {journal} {J.
  Chem. Phys.}\ }\textbf {\bibinfo {volume} {122}},\ \bibinfo {pages} {044106}
  (\bibinfo {year} {2005})}\BibitemShut {NoStop}%
\bibitem [{\citenamefont {Van\'{\i}\v{c}ek}\ \emph {et~al.}(2005)\citenamefont
  {Van\'{\i}\v{c}ek}, \citenamefont {Miller}, \citenamefont {Castillo},\ and\
  \citenamefont {Aoiz}}]{Vanicek2005}%
  \BibitemOpen
  \bibfield  {author} {\bibinfo {author} {\bibfnamefont {J.}~\bibnamefont
  {Van\'{\i}\v{c}ek}}, \bibinfo {author} {\bibfnamefont {W.~H.}\ \bibnamefont
  {Miller}}, \bibinfo {author} {\bibfnamefont {J.~F.}\ \bibnamefont
  {Castillo}}, \ and\ \bibinfo {author} {\bibfnamefont {F.~J.}\ \bibnamefont
  {Aoiz}},\ }\href@noop {} {\bibfield  {journal} {\bibinfo  {journal} {J. Chem.
  Phys.}\ }\textbf {\bibinfo {volume} {123}},\ \bibinfo {pages} {054108}
  (\bibinfo {year} {2005})}\BibitemShut {NoStop}%
\bibitem [{\citenamefont {Karandashev}\ \emph {et~al.}(2017)\citenamefont
  {Karandashev}, \citenamefont {Xu}, \citenamefont {Meuwly}, \citenamefont
  {Van\'{\i}\v{c}ek},\ and\ \citenamefont {Richardson}}]{Karandashev2017b}%
  \BibitemOpen
  \bibfield  {author} {\bibinfo {author} {\bibfnamefont {K.}~\bibnamefont
  {Karandashev}}, \bibinfo {author} {\bibfnamefont {Z.~H.}\ \bibnamefont {Xu}},
  \bibinfo {author} {\bibfnamefont {M.}~\bibnamefont {Meuwly}}, \bibinfo
  {author} {\bibfnamefont {J.}~\bibnamefont {Van\'{\i}\v{c}ek}}, \ and\
  \bibinfo {author} {\bibfnamefont {J.~O.}\ \bibnamefont {Richardson}},\
  }\href@noop {} {\bibfield  {journal} {\bibinfo  {journal} {Struct. Dyn.}\
  }\textbf {\bibinfo {volume} {4}},\ \bibinfo {pages} {061501} (\bibinfo {year}
  {2017})}\BibitemShut {NoStop}%
\bibitem [{\citenamefont {Karandashev}\ and\ \citenamefont
  {Van\'{\i}\v{c}ek}(2015)}]{Karandashev2015}%
  \BibitemOpen
  \bibfield  {author} {\bibinfo {author} {\bibfnamefont {K.}~\bibnamefont
  {Karandashev}}\ and\ \bibinfo {author} {\bibfnamefont {J.}~\bibnamefont
  {Van\'{\i}\v{c}ek}},\ }\href@noop {} {\bibfield  {journal} {\bibinfo
  {journal} {J. Chem. Phys.}\ }\textbf {\bibinfo {volume} {143}},\ \bibinfo
  {pages} {194104} (\bibinfo {year} {2015})}\BibitemShut {NoStop}%
\bibitem [{\citenamefont {Karandashev}\ and\ \citenamefont
  {Van\'{\i}\v{c}ek}(2017)}]{Karandashev2017a}%
  \BibitemOpen
  \bibfield  {author} {\bibinfo {author} {\bibfnamefont {K.}~\bibnamefont
  {Karandashev}}\ and\ \bibinfo {author} {\bibfnamefont {J.}~\bibnamefont
  {Van\'{\i}\v{c}ek}},\ }\href@noop {} {\bibfield  {journal} {\bibinfo
  {journal} {J. Chem. Phys.}\ }\textbf {\bibinfo {volume} {146}},\ \bibinfo
  {pages} {184102} (\bibinfo {year} {2017})}\BibitemShut {NoStop}%
\bibitem [{\citenamefont {Karandashev}(2017)}]{Karandashev2017c}%
  \BibitemOpen
  \bibfield  {author} {\bibinfo {author} {\bibfnamefont {K.}~\bibnamefont
  {Karandashev}},\ }\emph {\bibinfo {title} {Accelerating path integral
  evaluation of equilibrium and kinetic isotope effects}},\ \href@noop {}
  {Ph.D. thesis},\ \bibinfo  {school} {\'Ecole Polytechnique F\'ed\'erale de
  Lausanne} (\bibinfo {year} {2017})\BibitemShut {NoStop}%
\bibitem [{\citenamefont {Chandler}(1978)}]{Chandler1978TST}%
  \BibitemOpen
  \bibfield  {author} {\bibinfo {author} {\bibfnamefont {D.}~\bibnamefont
  {Chandler}},\ }\href {\doibase 10.1063/1.436049} {\bibfield  {journal}
  {\bibinfo  {journal} {J.~Chem. Phys.}\ }\textbf {\bibinfo {volume} {68}},\
  \bibinfo {pages} {2959} (\bibinfo {year} {1978})}\BibitemShut {NoStop}%
\bibitem [{\citenamefont {Mills}\ \emph {et~al.}(1997)\citenamefont {Mills},
  \citenamefont {Schenter}, \citenamefont {Makarov},\ and\ \citenamefont
  {J{\'o}nsson}}]{Mills1997QTST}%
  \BibitemOpen
  \bibfield  {author} {\bibinfo {author} {\bibfnamefont {G.}~\bibnamefont
  {Mills}}, \bibinfo {author} {\bibfnamefont {G.~K.}\ \bibnamefont {Schenter}},
  \bibinfo {author} {\bibfnamefont {D.~E.}\ \bibnamefont {Makarov}}, \ and\
  \bibinfo {author} {\bibfnamefont {H.}~\bibnamefont {J{\'o}nsson}},\ }\href
  {\doibase 10.1016/S0009-2614(97)00886-5} {\bibfield  {journal} {\bibinfo
  {journal} {Chem. Phys. Lett.}\ }\textbf {\bibinfo {volume} {278}},\ \bibinfo
  {pages} {91} (\bibinfo {year} {1997})}\BibitemShut {NoStop}%
\bibitem [{\citenamefont {Hele}\ and\ \citenamefont
  {Althorpe}(2013)}]{Hele2013}%
  \BibitemOpen
  \bibfield  {author} {\bibinfo {author} {\bibfnamefont {T.~J.~H.}\
  \bibnamefont {Hele}}\ and\ \bibinfo {author} {\bibfnamefont {S.~C.}\
  \bibnamefont {Althorpe}},\ }\href@noop {} {\bibfield  {journal} {\bibinfo
  {journal} {J. Chem. Phys.}\ }\textbf {\bibinfo {volume} {138}},\ \bibinfo
  {pages} {084108} (\bibinfo {year} {2013})}\BibitemShut {NoStop}%
\bibitem [{\citenamefont {Althorpe}\ and\ \citenamefont
  {Hele}(2013)}]{Althorpe2013}%
  \BibitemOpen
  \bibfield  {author} {\bibinfo {author} {\bibfnamefont {S.~C.}\ \bibnamefont
  {Althorpe}}\ and\ \bibinfo {author} {\bibfnamefont {T.~J.~H.}\ \bibnamefont
  {Hele}},\ }\href@noop {} {\bibfield  {journal} {\bibinfo  {journal} {J. Chem.
  Phys.}\ }\textbf {\bibinfo {volume} {139}},\ \bibinfo {pages} {084115}
  (\bibinfo {year} {2013})}\BibitemShut {NoStop}%
\bibitem [{\citenamefont {Voth}, \citenamefont {Chandler},\ and\ \citenamefont
  {Miller}(1989)}]{Voth+Chandler+Miller}%
  \BibitemOpen
  \bibfield  {author} {\bibinfo {author} {\bibfnamefont {G.~A.}\ \bibnamefont
  {Voth}}, \bibinfo {author} {\bibfnamefont {D.}~\bibnamefont {Chandler}}, \
  and\ \bibinfo {author} {\bibfnamefont {W.~H.}\ \bibnamefont {Miller}},\
  }\href {\doibase 10.1063/1.457242} {\bibfield  {journal} {\bibinfo  {journal}
  {J.~Chem. Phys.}\ }\textbf {\bibinfo {volume} {91}},\ \bibinfo {pages} {7749}
  (\bibinfo {year} {1989})}\BibitemShut {NoStop}%
\bibitem [{\citenamefont {Makarov}\ and\ \citenamefont
  {Topaler}(1995)}]{Makarov1995QTST}%
  \BibitemOpen
  \bibfield  {author} {\bibinfo {author} {\bibfnamefont {D.~E.}\ \bibnamefont
  {Makarov}}\ and\ \bibinfo {author} {\bibfnamefont {M.}~\bibnamefont
  {Topaler}},\ }\href {\doibase 10.1103/PhysRevE.52.178} {\bibfield  {journal}
  {\bibinfo  {journal} {Phys. Rev. E}\ }\textbf {\bibinfo {volume} {52}},\
  \bibinfo {pages} {178} (\bibinfo {year} {1995})}\BibitemShut {NoStop}%
\bibitem [{\citenamefont {Aieta}\ and\ \citenamefont
  {Ceotto}(2017)}]{Aieta2017}%
  \BibitemOpen
  \bibfield  {author} {\bibinfo {author} {\bibfnamefont {C.}~\bibnamefont
  {Aieta}}\ and\ \bibinfo {author} {\bibfnamefont {M.}~\bibnamefont {Ceotto}},\
  }\href@noop {} {\bibfield  {journal} {\bibinfo  {journal} {J. Chem. Phys.}\
  }\textbf {\bibinfo {volume} {146}},\ \bibinfo {pages} {214115} (\bibinfo
  {year} {2017})}\BibitemShut {NoStop}%
\bibitem [{\citenamefont {Ceotto}, \citenamefont {Yang},\ and\ \citenamefont
  {Miller}(2005)}]{Ceotto2005}%
  \BibitemOpen
  \bibfield  {author} {\bibinfo {author} {\bibfnamefont {M.}~\bibnamefont
  {Ceotto}}, \bibinfo {author} {\bibfnamefont {S.}~\bibnamefont {Yang}}, \ and\
  \bibinfo {author} {\bibfnamefont {W.~H.}\ \bibnamefont {Miller}},\
  }\href@noop {} {\bibfield  {journal} {\bibinfo  {journal} {J. Chem. Phys.}\
  }\textbf {\bibinfo {volume} {122}},\ \bibinfo {pages} {044109} (\bibinfo
  {year} {2005})}\BibitemShut {NoStop}%
\bibitem [{\citenamefont {Yang}, \citenamefont {Yamamoto},\ and\ \citenamefont
  {Miller}(2006)}]{Yang2006}%
  \BibitemOpen
  \bibfield  {author} {\bibinfo {author} {\bibfnamefont {S.}~\bibnamefont
  {Yang}}, \bibinfo {author} {\bibfnamefont {T.}~\bibnamefont {Yamamoto}}, \
  and\ \bibinfo {author} {\bibfnamefont {W.~H.}\ \bibnamefont {Miller}},\
  }\href@noop {} {\bibfield  {journal} {\bibinfo  {journal} {J. Chem. Phys.}\
  }\textbf {\bibinfo {volume} {124}},\ \bibinfo {pages} {084102} (\bibinfo
  {year} {2006})}\BibitemShut {NoStop}%
\bibitem [{\citenamefont {Predescu}(2004)}]{Predescu2004}%
  \BibitemOpen
  \bibfield  {author} {\bibinfo {author} {\bibfnamefont {C.}~\bibnamefont
  {Predescu}},\ }\href@noop {} {\bibfield  {journal} {\bibinfo  {journal}
  {Phys. Rev. E}\ }\textbf {\bibinfo {volume} {70}},\ \bibinfo {pages} {066705}
  (\bibinfo {year} {2004})}\BibitemShut {NoStop}%
\bibitem [{Note1()}]{Note1}%
  \BibitemOpen
  \bibinfo {note} {We were unable to locate an asymmetric system with split
  saddle points of $C_\protect \mathrm {ff}(0)$.}\BibitemShut {Stop}%
\bibitem [{\citenamefont {Huo}, \citenamefont {Miller~III},\ and\ \citenamefont
  {Coker}(2013)}]{Huo2013PLDM}%
  \BibitemOpen
  \bibfield  {author} {\bibinfo {author} {\bibfnamefont {P.}~\bibnamefont
  {Huo}}, \bibinfo {author} {\bibfnamefont {T.~F.}\ \bibnamefont {Miller~III}},
  \ and\ \bibinfo {author} {\bibfnamefont {D.~F.}\ \bibnamefont {Coker}},\
  }\href {\doibase 10.1063/1.4826163} {\bibfield  {journal} {\bibinfo
  {journal} {J.~Chem. Phys.}\ }\textbf {\bibinfo {volume} {139}},\ \bibinfo
  {pages} {151103} (\bibinfo {year} {2013})}\BibitemShut {NoStop}%
\bibitem [{\citenamefont {Richardson}\ and\ \citenamefont
  {Thoss}(2014)}]{nonoscillatory}%
  \BibitemOpen
  \bibfield  {author} {\bibinfo {author} {\bibfnamefont {J.~O.}\ \bibnamefont
  {Richardson}}\ and\ \bibinfo {author} {\bibfnamefont {M.}~\bibnamefont
  {Thoss}},\ }\href {\doibase 10.1063/1.4892865} {\bibfield  {journal}
  {\bibinfo  {journal} {J.~Chem. Phys.}\ }\textbf {\bibinfo {volume} {141}},\
  \bibinfo {pages} {074106} (\bibinfo {year} {2014})}\BibitemShut {NoStop}%
\bibitem [{\citenamefont {Miller}(1971)}]{Miller1971density}%
  \BibitemOpen
  \bibfield  {author} {\bibinfo {author} {\bibfnamefont {W.~H.}\ \bibnamefont
  {Miller}},\ }\href {\doibase 10.1063/1.1676560} {\bibfield  {journal}
  {\bibinfo  {journal} {J.~Chem. Phys.}\ }\textbf {\bibinfo {volume} {55}},\
  \bibinfo {pages} {3146} (\bibinfo {year} {1971})}\BibitemShut {NoStop}%
\bibitem [{\citenamefont {Gutzwiller}(1990)}]{Gutzwiller1990}%
  \BibitemOpen
  \bibfield  {author} {\bibinfo {author} {\bibfnamefont {M.~C.}\ \bibnamefont
  {Gutzwiller}},\ }\href@noop {} {\emph {\bibinfo {title} {Chaos in Classical
  and Quantum Mechanics}}}\ (\bibinfo  {publisher} {Springer-Verlag: New
  York},\ \bibinfo {year} {1990})\BibitemShut {NoStop}%
\bibitem [{\citenamefont {Bender}\ and\ \citenamefont
  {Orszag}(1978)}]{BenderBook}%
  \BibitemOpen
  \bibfield  {author} {\bibinfo {author} {\bibfnamefont {C.~M.}\ \bibnamefont
  {Bender}}\ and\ \bibinfo {author} {\bibfnamefont {S.~A.}\ \bibnamefont
  {Orszag}},\ }\href@noop {} {\emph {\bibinfo {title} {Advanced Mathematical
  Methods for Scientists and Engineers}}}\ (\bibinfo  {publisher}
  {McGraw-Hill},\ \bibinfo {address} {New York},\ \bibinfo {year}
  {1978})\BibitemShut {NoStop}%
\bibitem [{\citenamefont {Richardson}, \citenamefont {Bauer},\ and\
  \citenamefont {Thoss}(2015)}]{Richardson2015}%
  \BibitemOpen
  \bibfield  {author} {\bibinfo {author} {\bibfnamefont {J.~O.}\ \bibnamefont
  {Richardson}}, \bibinfo {author} {\bibfnamefont {R.}~\bibnamefont {Bauer}}, \
  and\ \bibinfo {author} {\bibfnamefont {M.}~\bibnamefont {Thoss}},\
  }\href@noop {} {\bibfield  {journal} {\bibinfo  {journal} {J. Chem. Phys.}\
  }\textbf {\bibinfo {volume} {143}},\ \bibinfo {pages} {134115} (\bibinfo
  {year} {2015})}\BibitemShut {NoStop}%
\bibitem [{\citenamefont {Wolynes}(1987)}]{Wolynes1987nonadiabatic}%
  \BibitemOpen
  \bibfield  {author} {\bibinfo {author} {\bibfnamefont {P.~G.}\ \bibnamefont
  {Wolynes}},\ }\href {\doibase 10.1063/1.453440} {\bibfield  {journal}
  {\bibinfo  {journal} {J.~Chem. Phys.}\ }\textbf {\bibinfo {volume} {87}},\
  \bibinfo {pages} {6559} (\bibinfo {year} {1987})}\BibitemShut {NoStop}%
\bibitem [{\citenamefont {Thapa}, \citenamefont {Fang},\ and\ \citenamefont
  {Richardson}(2019)}]{GRQTST}%
  \BibitemOpen
  \bibfield  {author} {\bibinfo {author} {\bibfnamefont {M.~J.}\ \bibnamefont
  {Thapa}}, \bibinfo {author} {\bibfnamefont {W.}~\bibnamefont {Fang}}, \ and\
  \bibinfo {author} {\bibfnamefont {J.~O.}\ \bibnamefont {Richardson}},\ }\href
  {\doibase 10.1063/1.5081108} {\bibfield  {journal} {\bibinfo  {journal} {J.
  Chem. Phys.}\ }\textbf {\bibinfo {volume} {150}},\ \bibinfo {pages} {104107}
  (\bibinfo {year} {2019})}\BibitemShut {NoStop}%
\bibitem [{Note2()}]{Note2}%
  \BibitemOpen
  \bibinfo {note} {The error in the QI rates is slightly larger than those
  reported in Ref.~\protect \rev@citealpnum {Miller2003} because they used an
  \protect \emph {ad hoc} correction to the energy variance for their
  calculations, but this does not affect our conclusions.}\BibitemShut {Stop}%
\bibitem [{\citenamefont {Richardson}(2016{\natexlab{b}})}]{Faraday}%
  \BibitemOpen
  \bibfield  {author} {\bibinfo {author} {\bibfnamefont {J.~O.}\ \bibnamefont
  {Richardson}},\ }\href {\doibase 10.1039/C6FD00119J} {\bibfield  {journal}
  {\bibinfo  {journal} {Faraday Discuss.}\ }\textbf {\bibinfo {volume} {195}},\
  \bibinfo {pages} {49} (\bibinfo {year} {2016}{\natexlab{b}})}\BibitemShut
  {NoStop}%
\bibitem [{\citenamefont {Habershon}\ \emph {et~al.}(2013)\citenamefont
  {Habershon}, \citenamefont {Manolopoulos}, \citenamefont {Markland},\ and\
  \citenamefont {Miller~III}}]{Habershon2013RPMDreview}%
  \BibitemOpen
  \bibfield  {author} {\bibinfo {author} {\bibfnamefont {S.}~\bibnamefont
  {Habershon}}, \bibinfo {author} {\bibfnamefont {D.~E.}\ \bibnamefont
  {Manolopoulos}}, \bibinfo {author} {\bibfnamefont {T.~E.}\ \bibnamefont
  {Markland}}, \ and\ \bibinfo {author} {\bibfnamefont {T.~F.}\ \bibnamefont
  {Miller~III}},\ }\href {\doibase 10.1146/annurev-physchem-040412-110122}
  {\bibfield  {journal} {\bibinfo  {journal} {Annu. Rev. Phys. Chem.}\ }\textbf
  {\bibinfo {volume} {64}},\ \bibinfo {pages} {387} (\bibinfo {year}
  {2013})}\BibitemShut {NoStop}%
\bibitem [{\citenamefont {Feynman}\ and\ \citenamefont
  {Hibbs}(1965)}]{Feynman1965}%
  \BibitemOpen
  \bibfield  {author} {\bibinfo {author} {\bibfnamefont {R.~P.}\ \bibnamefont
  {Feynman}}\ and\ \bibinfo {author} {\bibfnamefont {A.~R.}\ \bibnamefont
  {Hibbs}},\ }\href@noop {} {\emph {\bibinfo {title} {Quantum Mechanics and
  Path Integrals}}}\ (\bibinfo  {publisher} {McGraw-Hill: New York},\ \bibinfo
  {year} {1965})\BibitemShut {NoStop}%
\bibitem [{\citenamefont {Gradshteyn}\ and\ \citenamefont
  {Ryzhik}(2000)}]{Gradshteyn}%
  \BibitemOpen
  \bibfield  {author} {\bibinfo {author} {\bibfnamefont {I.~S.}\ \bibnamefont
  {Gradshteyn}}\ and\ \bibinfo {author} {\bibfnamefont {I.~M.}\ \bibnamefont
  {Ryzhik}},\ }\href@noop {} {\emph {\bibinfo {title} {Tables of Integrals,
  Series and Products}}},\ \bibinfo {edition} {6th}\ ed.\ (\bibinfo
  {publisher} {Academic Press},\ \bibinfo {address} {San Diego},\ \bibinfo
  {year} {2000})\BibitemShut {NoStop}%
\bibitem [{\citenamefont {Miller}\ \emph
  {et~al.}(2003{\natexlab{b}})\citenamefont {Miller}, \citenamefont {Zhao},
  \citenamefont {Ceotto},\ and\ \citenamefont {Yang}}]{Miller2003QI}%
  \BibitemOpen
  \bibfield  {author} {\bibinfo {author} {\bibfnamefont {W.~H.}\ \bibnamefont
  {Miller}}, \bibinfo {author} {\bibfnamefont {Y.}~\bibnamefont {Zhao}},
  \bibinfo {author} {\bibfnamefont {M.}~\bibnamefont {Ceotto}}, \ and\ \bibinfo
  {author} {\bibfnamefont {S.}~\bibnamefont {Yang}},\ }\href {\doibase
  10.1063/1.1580110} {\bibfield  {journal} {\bibinfo  {journal} {J.~Chem.
  Phys.}\ }\textbf {\bibinfo {volume} {119}},\ \bibinfo {pages} {1329}
  (\bibinfo {year} {2003}{\natexlab{b}})}\BibitemShut {NoStop}%
\bibitem [{\citenamefont {Hansen}\ and\ \citenamefont
  {Andersen}(1996)}]{Hansen1996QTST}%
  \BibitemOpen
  \bibfield  {author} {\bibinfo {author} {\bibfnamefont {N.~F.}\ \bibnamefont
  {Hansen}}\ and\ \bibinfo {author} {\bibfnamefont {H.~C.}\ \bibnamefont
  {Andersen}},\ }\href@noop {} {\bibfield  {journal} {\bibinfo  {journal}
  {J.~Phys. Chem.}\ }\textbf {\bibinfo {volume} {100}},\ \bibinfo {pages}
  {1137} (\bibinfo {year} {1996})}\BibitemShut {NoStop}%
\bibitem [{\citenamefont {Eckart}(1930)}]{Eckart1930}%
  \BibitemOpen
  \bibfield  {author} {\bibinfo {author} {\bibfnamefont {C.}~\bibnamefont
  {Eckart}},\ }\href@noop {} {\bibfield  {journal} {\bibinfo  {journal} {Phys.
  Rev.}\ }\textbf {\bibinfo {volume} {35}},\ \bibinfo {pages} {1303} (\bibinfo
  {year} {1930})}\BibitemShut {NoStop}%
\bibitem [{\citenamefont {Klein}(1933)}]{Klein}%
  \BibitemOpen
  \bibfield  {author} {\bibinfo {author} {\bibfnamefont {F.}~\bibnamefont
  {Klein}},\ }\href@noop {} {\emph {\bibinfo {title} {Vorlesungen {\"u}ber die
  hypergeometrische Funktion}}}\ (\bibinfo  {publisher} {Springer-Verlag},\
  \bibinfo {year} {1933})\ pp.\ \bibinfo {pages} {3--4}\BibitemShut {NoStop}%
\bibitem [{\citenamefont {Forsyth}(1948)}]{Forsyth}%
  \BibitemOpen
  \bibfield  {author} {\bibinfo {author} {\bibfnamefont {A.~R.}\ \bibnamefont
  {Forsyth}},\ }\href@noop {} {\emph {\bibinfo {title} {A treatise on
  differential equations}}},\ \bibinfo {edition} {6th}\ ed.\ (\bibinfo
  {publisher} {Macmillan},\ \bibinfo {year} {1948})\ pp.\ \bibinfo {pages}
  {214--215}\BibitemShut {NoStop}%
\bibitem [{\citenamefont {Abramowitz}\ and\ \citenamefont
  {Stegun}(1972)}]{Abramowitz}%
  \BibitemOpen
  \bibfield  {author} {\bibinfo {author} {\bibfnamefont {M.}~\bibnamefont
  {Abramowitz}}\ and\ \bibinfo {author} {\bibfnamefont {I.~A.}\ \bibnamefont
  {Stegun}},\ }\href@noop {} {\emph {\bibinfo {title} {Handbook of Mathematical
  Functions}}}\ (\bibinfo  {publisher} {U.S. Government Printing Office},\
  \bibinfo {year} {1972})\BibitemShut {NoStop}%
\end{thebibliography}%
\end{document}